\documentclass[
prd 
,showpacs ,amssymb,superscriptaddress,aps
]{revtex4}
\usepackage{graphicx}
\input{epsf}

\usepackage{amsmath,amssymb}
\usepackage{bm}
\usepackage{times}

\newcommand{\dalm}{\kern1pt\vbox{\hrule height 0.9pt\hbox{\vrule width 0.9pt
\hskip 2.5pt\vbox{\vskip 5.5pt}\hskip 3pt\vrule width 0.3pt}\hrule height 0.3pt}
\kern1pt}

\begin{document}



\title{Torsional oscillations of neutron stars with highly tangled magnetic fields}

\author{Hajime Sotani}
\email{hajime.sotani@nao.ac.jp}
\affiliation{Division of Theoretical Astronomy, National Astronomical Observatory of Japan, 2-21-1 Osawa, Mitaka, Tokyo 181-8588, Japan}


\date{\today}

\begin{abstract}
To determine the frequencies of magnetic oscillations in neutron stars with highly tangled magnetic fields, we derive the perturbation equations. We assume that the field strength of the global magnetic structure is so small that such fields are negligible compared with tangled fields, which may still be far from a realistic configuration. Then, we systematically examine the spectra of the magnetic oscillations, as varying the magnetic field strength and stellar mass. The frequencies without crust elasticity are completely proportional to the strength of the magnetic field, whose proportionality constant depends strongly on the stellar mass. On the other hand, the oscillation spectra with crust elasticity become more complicated, where the frequencies even for weak magnetic fields are different from the crustal torsional oscillations without magnetic fields. For discussing spectra, the critical field strength can play an important role, and it is determined in such a way that the shear velocity is equivalent to the Alfv\'{e}n velocity at the crust basis. Additionally, we find that the effect of the crust elasticity can be seen strongly in the fundamental oscillations with a lower harmonic index, $\ell$. Unlike the stellar models with a pure dipole magnetic field, we also find that the spectra with highly tangled magnetic fields become discrete, where one can expect many of the eigenfrequencies. Maybe these frequencies could be detected after the violent phenomena breaking the global magnetic field structure. 
\end{abstract}

\pacs{04.40.Dg, 97.10.Sj, 04.40.Nr}
%
\maketitle
\section{Introduction}
\label{sec:I}

Through the observations of pulsars, it is known that neutron stars generally have magnetic fields, whose typical strength is thought to be around $10^{12}-10^{13}$ Gauss, assuming that their spin periods would decay by the dipole radiations \cite{shapiro-teukolsky}. Additionally, the existence of the strongly magnetized neutron stars, the so-called magnetars \cite{DT1992}, is observationally suggested. Magnetars may sometimes be considered as a different family from the usual neutron stars. The magnetic field strength of magnetars becomes around $10^{14}-10^{15}$ Gauss. Unfortunately, the mechanism for how such a strong magnetic field is produced is still unknown. The magnetic fields outside neutron stars, including magnetars, can be dominated by dipole fields, but the magnetic distribution inside the star is also still unknown. The magnetic fields inside the star may eventually settle into dipole fields, but the magnetic configuration must be quite complicated at least just after the formation of neutron stars via supernova or merger of neutron stars. Under such a situation, where the magnetic fields are completely tangled inside the stars, the spectra of oscillations could be different from those expected in the cold neutron stars with quiet magnetic fields.

The observations of the spectra of stellar oscillations give us a good chance to extract the interior information of the neutron stars. Like seismology of the Earth and helioseismology of the Sun, asteroseismology of neutron stars is a valuable technique to obtain the interior information via the oscillation spectra, which could tell us the equation of state (EOS), stellar average density, stellar compactness, and so on \cite{AK1996,STM2001,SKH2004,SYMT2011,PA2012,DGKK2013}. Furthermore, via oscillation spectra from neutron stars, it might be possible to probe the gravitational theory itself in a strong gravitational field \cite{SK2004,Sotani2009,S2014a,S2014c}. The gravitational waves emitted from neutron stars are probably the most suitable candidates to adopt asteroseismology due to their strong penetration capability, although unfortunately they cannot be detected  yet. On the other hand, observational evidences of the oscillations of neutron stars already exists: the quasiperiodic oscillations radiated from soft-gamma repeaters.

Quasiperiodic oscillations are discovered in the afterglow of the giant flares observed in the soft-gamma repeaters \cite{I2005,SW2005,SW2006}. Since the soft-gamma repeaters are one of the candidates of magnetars, the discovered quasiperiodic oscillations are considered to be strongly associated with the oscillations of neutron stars. In order to theoretically explain the frequencies of quasiperiodic oscillations, many attempts have been made in terms of the crustal torsional oscillations and/or magnetic oscillations \cite{SA2007,Sotani2007,Sotani2008a,Sotani2008b,SK2009,CBK2009,CSF2009,PL2013,vHL2011,CK2011,vHL2012,GCFMS2012,GCFMS2013,PL2014}. From the point of view of asteroseismology, the possibilities for constraining the crust EOS using the observed frequencies have also been suggested  \cite{Sotani2011,SNIO2012,SNIO2013a,SNIO2013b,SIO2015b}. Meanwhile, through such attempts, the features of the magnetar oscillations have become increasingly understood. That is, the magnetic oscillations without the crust elasticity become continuum spectra, assuming a dipole magnetic field \cite{Sotani2008a,CBK2009,CSF2009,PL2013}, because the propagating time along the field lines inside the star is not a specific but a continuous quantity. Additionally, the excited oscillations inside the star with crust elasticity depend strongly on the strength of magnetic fields. The magnetic oscillations are only excited if the magnetic fields are strong enough, while the oscillations in the vicinity of the stellar surface become the crustal torsional oscillations if the magnetic fields are weak enough \cite{GCFMS2012}. In any case, the spectra of excited magnetic oscillations should depend on the magnetic configuration inside the star \cite{GCFMS2013}. Namely, since the magnetic fields just after the birth of a neutron star could be highly tangled, the spectra of the stellar oscillations might be different from those of the ``clean" magnetic configuration.

So far there have been very few examinations of the spectra analysis for the stellar models with highly tangled magnetic fields. At most, the stellar oscillations have been examined in simple toy models with constant density \cite{vHL2011,LvE2015}, where they consider the oscillations under the assumption of the uniform magnetic fields together with the tangled fields. In such studies, the authors point out the possibility that the existence of magnetic fields could modify the shear modulus inside the star, i.e., the introduction of an effective shear modulus. In this paper, deriving the perturbation equations in the realistic neutron star models, we focus especially on the highly tangled magnetic fields so that the global magnetic structure is negligible as an extreme case, and we systematically examine the magnetic oscillations in such stellar models. We remark that, since the actual field in magnetar interiors may have the global and tangled components, the limiting case of a purely tangled field considered in this paper may still be far from a realistic configuration, at least for expressing the cold neutron stars. In this examination, we consider the magnetic oscillations with and without crust elasticity. This is because the crust region does not form just after the production of neutron stars, and it could take time to appear \cite{Suwa2014}. We adopt geometric units, $c=G=1$, where $c$ and $G$ denote the speed of light and the gravitational constant, respectively, and the metric signature is $(-,+,+,+)$ in this paper.

\section{Linearized equations}
\label{sec:II}

\begin{figure}
\begin{center}
\includegraphics[scale=0.5]{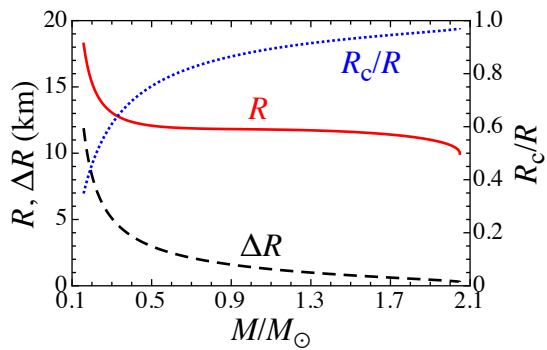} 
\end{center}
\caption{
The stellar radius, $R$, and the crust thickness, $\Delta R$, as a function of the stellar mass constructed with the SLy4 EOS. In addition, the ratio of the core radius to the stellar radius, $R_{\rm c}/R$, is also plotted with the dotted line.
}
\label{fig:M-RDR}
\end{figure}

In general, the magnetized neutron stars are deformed due to nonspherical magnetic pressure. However, the magnetic energy ($E_{\rm M}$) is much smaller than the gravitational binding energy ($E_{\rm G}$) even for the magnetars, where $E_{\rm M}/E_{\rm G}\sim 10^{-4}(B/10^{16} {\rm G})$. Thus, in this paper we neglect the deformation due to the existence of magnetic fields. Additionally, the stellar deformation may become significant only when the neutron star rotates very fast. In this paper, for simplicity, we also neglect the rotational effect, which leads to the consequence that the stellar configuration becomes spherically symmetric. The metric describing the spherically symmetric neutron stars is given by
\begin{equation}
 ds^2 = -e^{2\Phi}dt^2 + e^{2\Lambda}dr^2 + r^2(d \theta^2 + \sin^2\theta d \phi^2), \label{metric}
\end{equation}
where $\Phi$ and $\Lambda$ are the metric functions depending only on the radial coordinate $r$. In particular, $\Lambda$ is associated with the mass function $m(r)$ as $e^{-2\Lambda(r)}=1-2m(r)/r$. We remark that, from the metric form, the four-velocity of the equilibrium stellar model is written as $u^{\mu}=(e^{-\Phi},0,0,0)$. The stellar models are constructed by integrating the well-known Tolman-Oppenheimer-Volkoff equations together with the EOS of neutron star matter. In this paper, we adopt the EOS based on the Skyrme-type effective interaction, the so-called SLy4, which was derived by Douchin and  Haensel \cite{DH2001}. The density at the crust basis predicted from this EOS is $1.28 \times 10^{14}$ g/cm$^3$, while the maximum mass of the neutron star becomes $2.05M_\odot$, where the stellar radius is $10.0$ km. In Fig. \ref{fig:M-RDR}, we plot the stellar radius, $R$, and the crust thickness, $\Delta R$, as a function of the stellar mass. In the same figure, we also plot the ratio of the core radius, i.e., $R_{\rm c}\equiv R-\Delta R$, to the stellar radius with the dotted line. From this figure, one can see that the curst thickness is only less than 10 \% of the stellar radius for the neutron stars with a mass greater than $1.4M_\odot$.

On such an equilibrium configuration, we consider the axisymmetric axial perturbations, adopting the relativistic Cowling approximation. That is, we neglect the metric perturbations during the stellar oscillations. Since the axial perturbations do not involve the density variation, one can expect to determine the frequencies of the stellar oscillations with reasonable accuracy even for the relativistic Cowling approximation. The nonzero component of perturbed matter quantity in the axial perturbation is only $\delta u^{\phi}$, which is given by
\begin{equation}
  \delta u^\phi = e^{-\Phi}\partial_t {\cal Y}(t,r)b(\theta), 
\end{equation}
where $b(\theta)\equiv \sin^{-1}\theta\, \partial_\theta P_{\ell}(\cos\theta)$; $\partial_t$ and $\partial_\theta$ denote the partial derivative with respect to $t$ and $\theta$, respectively; ${\cal Y}(t,r)$ denotes the radial dependence of the angular displacement of matter element; and $P_{\ell}$ is the Legendre polynomial of order $\ell$.

The matter perturbations are described by the linearized equation of motion, i.e., Eq. (35) in \cite{Sotani2007}, while the perturbations of magnetic fields are subject to the linearized induction equation (37) in \cite{Sotani2007}. We remark that, assuming the ideal magnetohydrodynamics, the perturbations of magnetic fields can be written by the variable of matter perturbations together with the background quantities, as in Eq. (37) in \cite{Sotani2007}. Then, for a general magnetic distribution $H^\mu(r,\theta,\phi)$, the perturbation equation to determine the axisymmetric axial perturbation for the magnetized neutron stars becomes Eq. (51) in \cite{Sotani2007}, where $H^{\mu}$ is a normalized magnetic field given by $H^\mu\equiv B^\mu/\sqrt{4\pi}$ and the all perturbation variables are assumed to have a harmonic time dependence such as ${\cal Y}(t,r)=e^{i\omega t}{\cal Y}(r)$.

Now, we consider the decomposition of the magnetic fields $H^\mu$ as
\begin{equation}
  H^\mu = H^\mu_{\rm (G)} + H^\mu_{\rm (T)},
\end{equation}
where $H^\mu_{\rm (G)}$ and $H^\mu_{\rm (T)}$ denote the components of the global configuration and the intricately tangled structure of the magnetic fields inside the neutron stars, respectively. Here, we assume that the typical length scale of the tangled magnetic fields would be $\ell_{\rm T}$. That is, the small scale structure of the tangled magnetic fields should be taken into account, only if one focuses on the phenomena with a length scale smaller than $\ell_{\rm T}$. Otherwise, it is not necessary to care about the exact structure of the tangled magnetic fields, where only the strength of magnetic fields averaged in the volume $\ell_{\rm T}^3$ is a relevant quantity. Moreover, in order to examine the effects of the highly tangled magnetic fields on the stellar oscillations, we consider the situation that $H^\mu_{\rm (G)}\ll H^\mu_{\rm (T)}$ in this paper, i.e., $H^\mu \simeq H^\mu_{\rm (T)}$. These situations might be possible when the neutron star would be formed just after the supernovae or the merger of two neutron stars. Hereafter, we focus only on the phenomena with a length scale larger than $\ell_{\rm T}$, and we simply express $H^\mu_{\rm (T)}$ as $H^\mu$.

With respect to the highly tangled magnetic fields in the length scale larger than $\ell_{\rm T}$, it is natural to assume that $H^i$ does not have any correlations with $H^j$, $H^j_{\ ,k}$ for $i\ne j$, and $H^i_{\ ,k}$, and also that $H^i_{\ ,k}$ does not have any correlations with $H^j_{\ ,m}$ for $i\ne j$. Thus, in Eq. (51) in \cite{Sotani2007} one can set that, for $i\ne j$,
\begin{equation}
  H^iH^j = H^iH^j_{\ ,k} = H^iH^i_{\ ,k} = H^i_{\ ,k}H^j_{\ ,m}=0,  \label{eq:correlation}
\end{equation}
as in \cite{LvE2015}. As a result, one can obtain the linearized equation from Eq. (51) in \cite{Sotani2007}, such as
\begin{align}
 -\big[\varepsilon + p &+ H^rH_r + H^\theta H_\theta \big] \omega^2 e^{-2\Phi} {\cal Y} \nonumber \\
    =& \left[\mu + H^rH_r \right] e^{-2\Lambda}{\cal Y}''
     + \left[\left(\frac{4}{r} + \Phi' - \Lambda' \right)\mu + \mu' + \left(\Phi' + \frac{2}{r}\right)H^rH_r\right]e^{-2\Lambda}{\cal Y}'  \nonumber \\
    &- \frac{(\ell+2)(\ell-1)}{r^2}\left[\mu + H^\theta H_\theta\right]{\cal Y}
    + \Bigg[H^r_{\ ,\phi}H_{r,\phi} + H^{\theta}_{\ ,\phi}H_{\theta,\phi}\Bigg]\frac{1}{r^2 \sin^2\theta} {\cal Y}
    -\cot\theta(H^{\theta})^2{\cal Y}\frac{\partial_{\theta} b}{b}, \label{eq:perturbation0}
\end{align}
where $\mu$ is the shear modulus characterizing the elasticity of the neutron star crust, and the prime denotes the partial derivative with respect to the radial coordinate $r$.

Meanwhile, since one can derive the following relation from the Maxwell equations (Eq. (13) in \cite{Sotani2007}),
\begin{equation}
  -\cot\theta H^\theta = H^r_{\ ,r} + H^\theta_{\ ,\theta} + H^\phi_{\ ,\phi} + \left( \Lambda' + \frac{2}{r}\right)H^r,
\end{equation}
the last term in Eq. (\ref{eq:perturbation0}) is removed, using Eq. (\ref{eq:correlation}). Then, by setting that $H^rH_r=H^\theta H_\theta = H^\phi H_\phi={\cal H}^2/3$ for considering the oscillations whose wavelength is greater than $\ell_{\rm T}$, and assuming that the background magnetic field would be axisymmetric for simplicity, the linearized equation (\ref{eq:perturbation0}) becomes 
\begin{align}
 \left[\varepsilon + p + \frac{2{\cal H}^2}{3} \right] \omega^2 e^{-2\Phi} {\cal Y}
    +& \left[\mu + \frac{{\cal H}^2}{3} \right] e^{-2\Lambda}{\cal Y}''
     + \left[\left(\frac{4}{r} + \Phi' - \Lambda' \right)\mu + \mu' + \left(\Phi' + \frac{2}{r}\right)\frac{{\cal H}^2}{3}\right]e^{-2\Lambda}{\cal Y}'  \nonumber \\
    &- \frac{(\ell+2)(\ell-1)}{r^2}\left[\mu + \frac{{\cal H}^2}{3}\right]{\cal Y}, \label{eq:perturbation1}
\end{align}
where ${\cal H}$ is the local magnetic field strength defined by ${\cal H}=(H^r H_r + H^\theta H_\theta + H^\phi H_\phi)^{1/2}$. This is the linearized equation for axial perturbations in magnetized neutron stars with highly tangled magnetic fields, which is more general than the toy models adopted in \cite{vHL2011,LvE2015}. At last, with the appropriate boundary conditions, the problem to solve becomes the eigenvalue problem with the eigenfrequencies $\omega$.

The boundary conditions to solve the eigenvalue problem should be imposed at the stellar center ($r=0$) and at the surface ($r=R$), which are the regularity condition at $r=0$ and the zero-torque condition at $r=R$ \cite{ST1983}. 
In the vicinity of the stellar center, from Eq. (\ref{eq:perturbation1}), ${\cal Y}(r)$ can be expressed as ${\cal Y}\sim r^n$ with the positive constant $n$,
which leads to the boundary condition that $r{\cal Y}' = n{\cal Y}$. On the other hand, the boundary condition at the stellar surface becomes ${\cal Y}'=0$.
In addition to the boundary conditions, we also impose the junction condition at the interface between the stellar core and crust region, such as a zero-traction condition \cite{Sotani2007,ST1983}. In concrete terms, the junction condition can be written as $\left(3\mu + {\cal H}^2\right){\cal Y}'_{(+)} = {\cal H}^2{\cal Y}'_{(-)}$, where the left- and right-hand sides correspond to the values at just outside and inside the crust basis, respectively.

\begin{figure}
\begin{center}
\includegraphics[scale=0.5]{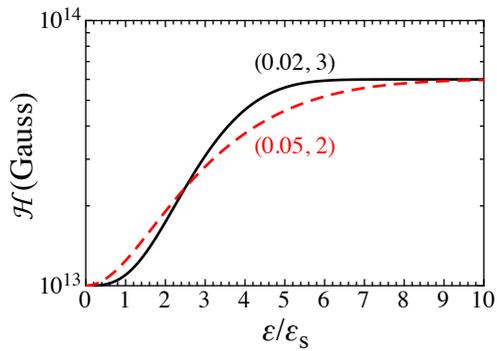} 
\end{center}
\caption{
Strength distribution of magnetic fields given by Eq. (\ref{eq:He}) with ${\cal H}_{\rm surf}=10^{13}$ Gauss. The solid and broken lines correspond to the cases with $(\beta,\gamma)=(0.02,3)$ and $(0.05,2)$, respectively.  
}
\label{fig:He}
\end{figure}

\begin{figure}
\begin{center}
\includegraphics[scale=0.5]{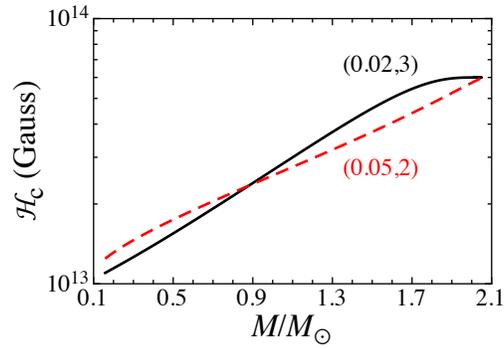} 
\end{center}
\caption{
With the strength distribution shown in Fig. \ref{fig:He}, the strength of magnetic fields at the stellar center, ${\cal H}_{\rm c}$, is plotted as a function of the stellar mass, where the solid and broken lines correspond to the cases with $(\beta,\gamma)=(0.02,3)$ and $(0.05,2)$, respectively.
}
\label{fig:MHc}
\end{figure}

In order to integrate Eq. (\ref{eq:perturbation1}), one has to prepare the strength distribution of ${\cal H}$, which is still unknown. So, as highly tangled magnetic fields, we adopt the density-dependent strength distribution of magnetic fields proposed in \cite{BCP1997}, i.e.,
\begin{equation}
  {\cal H}(\varepsilon/\varepsilon_s) = {\cal H}_{\rm surf} + {\cal H}_0\left[1-\exp\left\{-\beta(\varepsilon/\varepsilon_s)^\gamma\right\}\right], 
     \label{eq:He}
\end{equation}
where $\varepsilon_s$ denotes the saturation density, i.e., $\varepsilon_s=2.68\times 10^{14}$ g/cm$^3$, while ${\cal H}_{\rm surf}$ and ${\cal H}_0$ correspond to the magnetic field strength at the stellar surface and that for large density region. In this paper, we consider ${\cal H}_{\rm surf}$ as a free parameter, while ${\cal H}_0$ is chosen to be ${\cal H}_0=5\times {\cal H}_{\rm surf}$, which is a typical field strength at the stellar center for a dipole magnetic field \cite{ST2015}.  The remaining parameters, $\beta$ and $\gamma$, determine the magnetic structure inside the star. We especially examine the stellar oscillations with $(\beta,\gamma)=(0.02,3)$ and $(0.05,2)$, as in \cite{CCM2014}. In Fig. \ref{fig:He}, we show the strength distribution of magnetic fields given by Eq. (\ref{eq:He}) with ${\cal H}_{\rm surf}=10^{13}$ Gauss, where the solid and broken lines correspond to the cases of $(\beta,\gamma)=(0.02,3)$ and $(0.05,2)$, respectively. With such a strength distribution of magnetic fields, i.e., ${\cal H}_{\rm surf}=10^{13}$ Gauss, one can get the relationship between the expected strength at the stellar center, ${\cal H}_{\rm c}$, and the stellar mass, as shown in Fig. \ref{fig:MHc}.

Furthermore, we adopt the often-used shear modulus $\mu$ in the zero temperature limit, which is derived in \cite{OI1990,SHOII1991}, i.e.,
\begin{equation}
  \mu = 0.1194\frac{n_i(Ze)^2}{a}. \label{eq:shear}
\end{equation}
In the formula, $n_i$, $Z$, and $a$ denote the ion number density, the charge number of the ion, and the radius of a Wigner-Seitz cell, respectively. We remark that the shear modulus [Eq. (\ref{eq:shear})] is derived on the assumption that the nuclei form a body center cubic lattice due to the Coulomb interaction in the uniform distribution of electrons, which is averaged over all directions.

\section{Magnetic oscillations without crust elasticity}
\label{sec:III}

First, we consider the stellar oscillations without the effect of crust elasticity by setting $\mu=0$. The newly born neutron star, just after the supernova explosion or the merger of binary neutron stars, may not have a crust region, because the temperature is too high to form the crystallization of crust. In this case, the restoring force of the torsional oscillations inside the neutron stars is only the magnetic tension, which excites the magnetic oscillations. Thus, one can expect that the frequencies would be proportional to the strength of the magnetic fields.

\begin{figure*}
\begin{center}
\begin{tabular}{cc}
\includegraphics[scale=0.5]{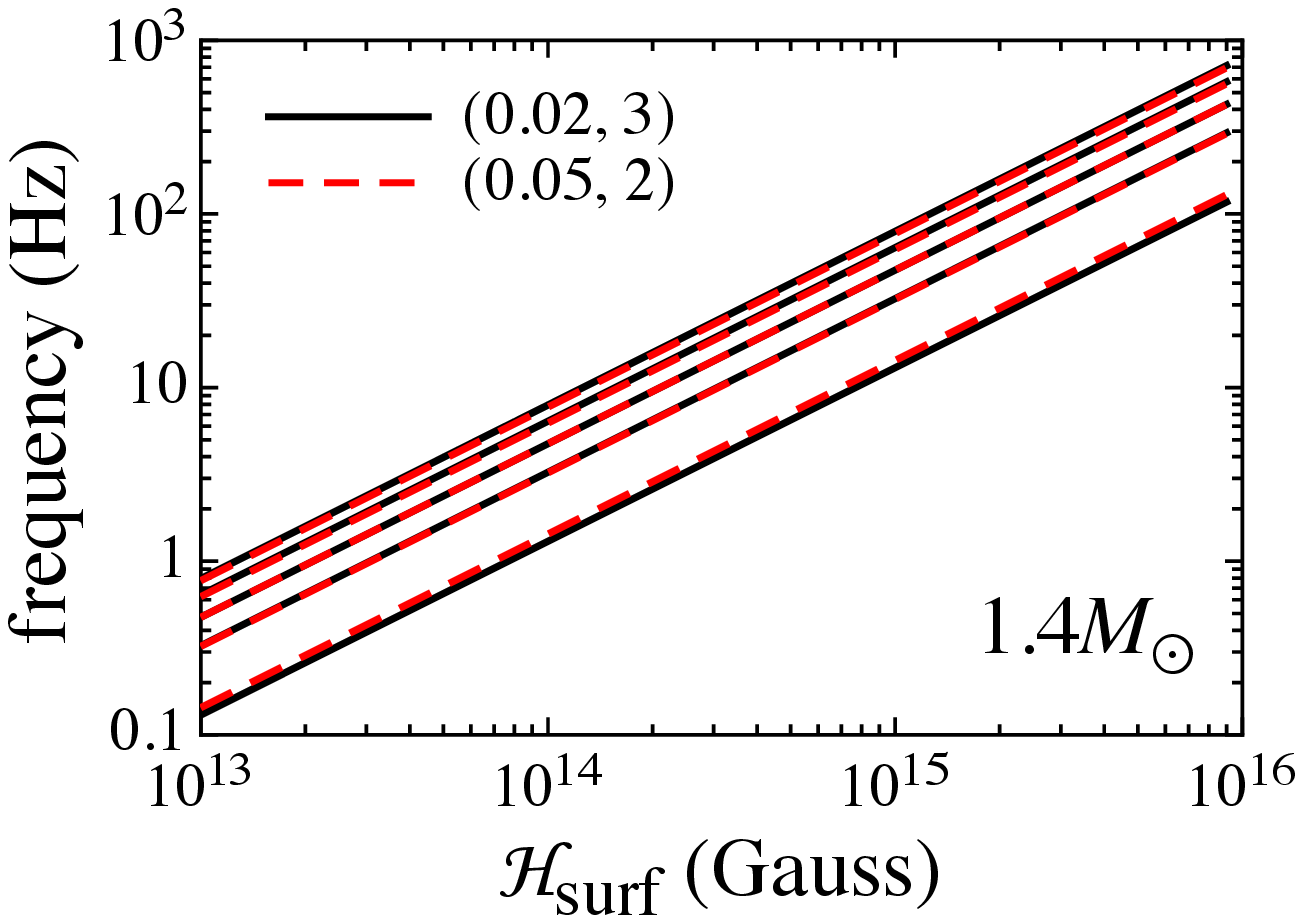} &
\includegraphics[scale=0.5]{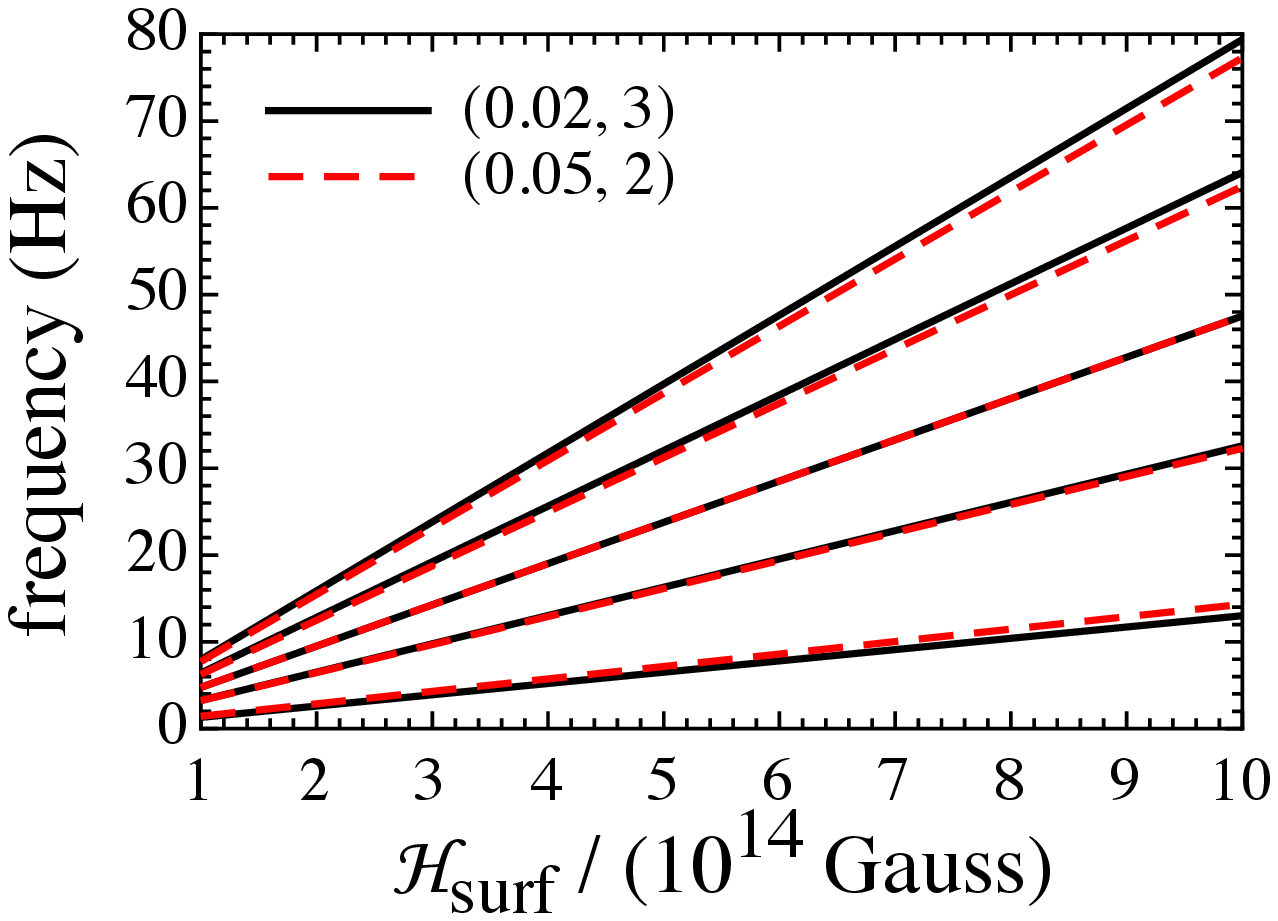}
\end{tabular}
\end{center}
\caption{
The lowest five frequencies of the $\ell=2$ torsional oscillations in the neutron star with highly tangled magnetic fields for the stellar model with $M=1.4M_\odot$, as a function of the strength of magnetic fields at the stellar surface, ${\cal H}_{\rm surf}$. In the figure, the solid and broken lines correspond to the frequencies obtained from the strength distribution of magnetic fields with $(\beta,\gamma)=(0.02,3)$ and $(0.05,2)$, respectively. The right panel is the magnified figure of the left panel.
}
\label{fig:Hsl2-M14}
\end{figure*}

\begin{figure*}
\begin{center}
\begin{tabular}{cc}
\includegraphics[scale=0.5]{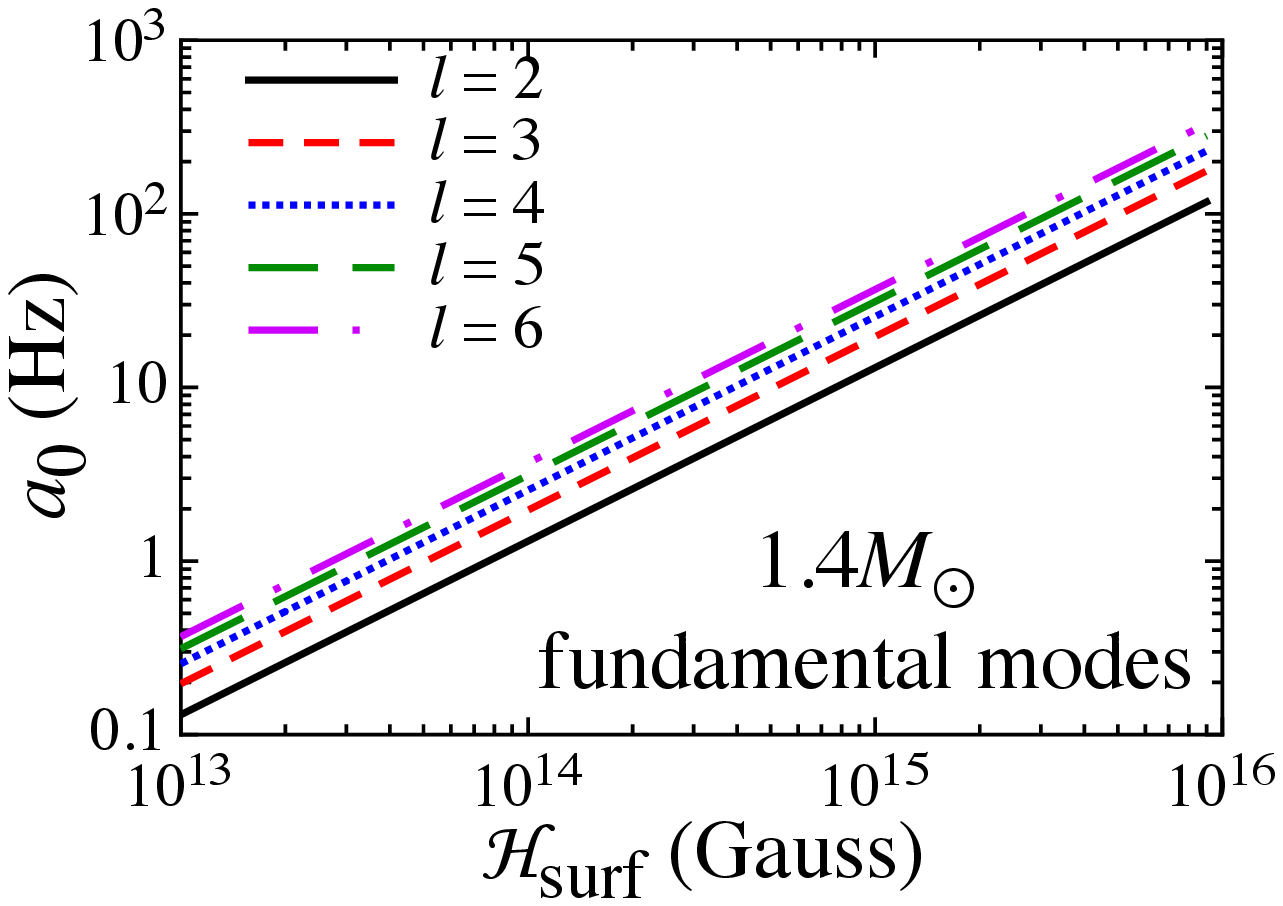} &
\includegraphics[scale=0.5]{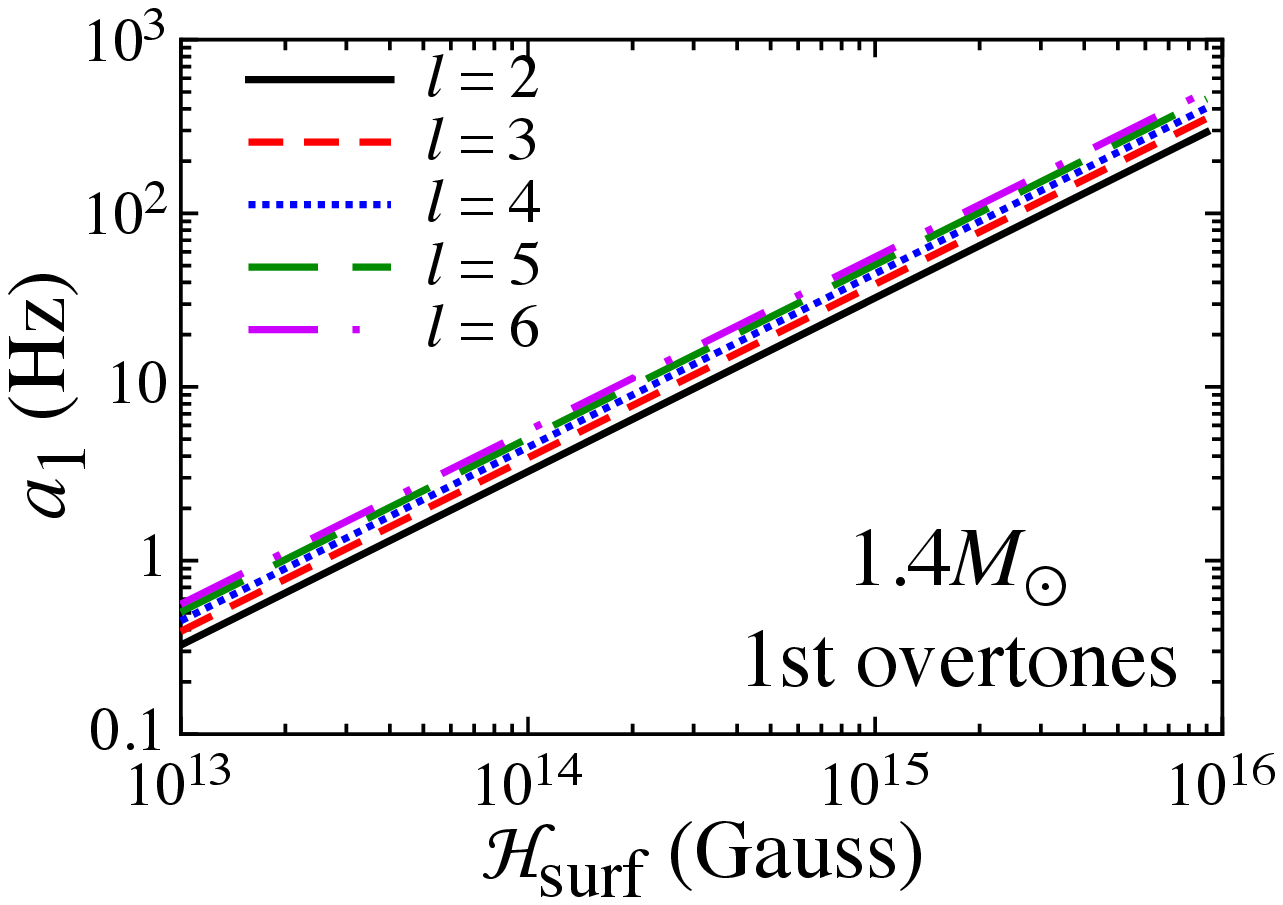}
\end{tabular}
\end{center}
\caption{
The frequencies of the fundamental oscillations, $a_0$, in the left panel and the 1st overtones, $a_1$, in the right panel for the $\ell = 2$, $3$, $4$, $5$, and $6$ oscillations in the stellar model with $M=1.4M_\odot$, as a function of ${\cal H}_{\rm surf}$, where the strength distribution of magnetic fields is assumed to be $(\beta,\gamma) = (0.02, 3)$.
}
\label{fig:Hsll-M14}
\end{figure*}

In Fig. \ref{fig:Hsl2-M14}, the calculated frequencies of the $\ell=2$ oscillations are shown as a function of the strength of the magnetic fields at the stellar surface, ${\cal H}_{\rm surf}$, for the stellar model with $M=1.4M_\odot$, where the solid and broken lines correspond to the results for the strength distribution of the magnetic fields with $(\beta,\gamma)=(0.02,3)$ and $(0.05,2)$, respectively. In the figure, we show the lowest five frequencies, i.e., the fundamental oscillations, 1st, 2nd, 3rd, and 4th overtones, while the right panel corresponds to the magnified figure of the left panel. In Fig. \ref{fig:Hsll-M14}, we also show the frequencies of the fundamental oscillations, $a_0$, in the left panel and the 1st overtones, $a_1$, in the right panel for the $\ell=2$, $3$, $4$, $5$, and $6$ oscillations in the stellar model with $M=1.4M_\odot$, as a function of ${\cal H}_{\rm surf}$, where the strength distribution of magnetic fields is assumed to be $(\beta,\gamma) = (0.02, 3)$. From both Figs. \ref{fig:Hsl2-M14} and \ref{fig:Hsll-M14}, as expected, one can observe that the frequencies of the magnetic oscillations inside the neutron stars are proportional to the strength of the magnetic fields, such as
\begin{equation}
   {}_\ell a_n = {}_\ell c_n \times\frac{{\cal H}_{\rm surf}}{10^{13}\, {\rm Gauss}},  \label{eq:lan}
\end{equation}
where ${}_\ell a_n$ denotes the frequencies of the $\ell$th magnetic oscillations with the number of radial nodes, $n$, while ${}_\ell c_n$ is a proportionality constant depending on the stellar model and the strength distribution of the magnetic fields.

We remark that, among many axial type oscillations, the $\ell=2$ fundamental oscillation, i.e., ${}_2 a_0$, is the lowest frequency theoretically expected. Thus, there are many eigenfrequencies above the line for ${}_2 a_0$ in Fig. \ref{fig:Hsl2-M14}, depending on the values of $\ell$ and $n$, but one cannot expect the existence of the eigenfrequencies below the line for ${}_2 a_0$.

\begin{figure*}
\begin{center}
\begin{tabular}{ccc}
\includegraphics[scale=0.43]{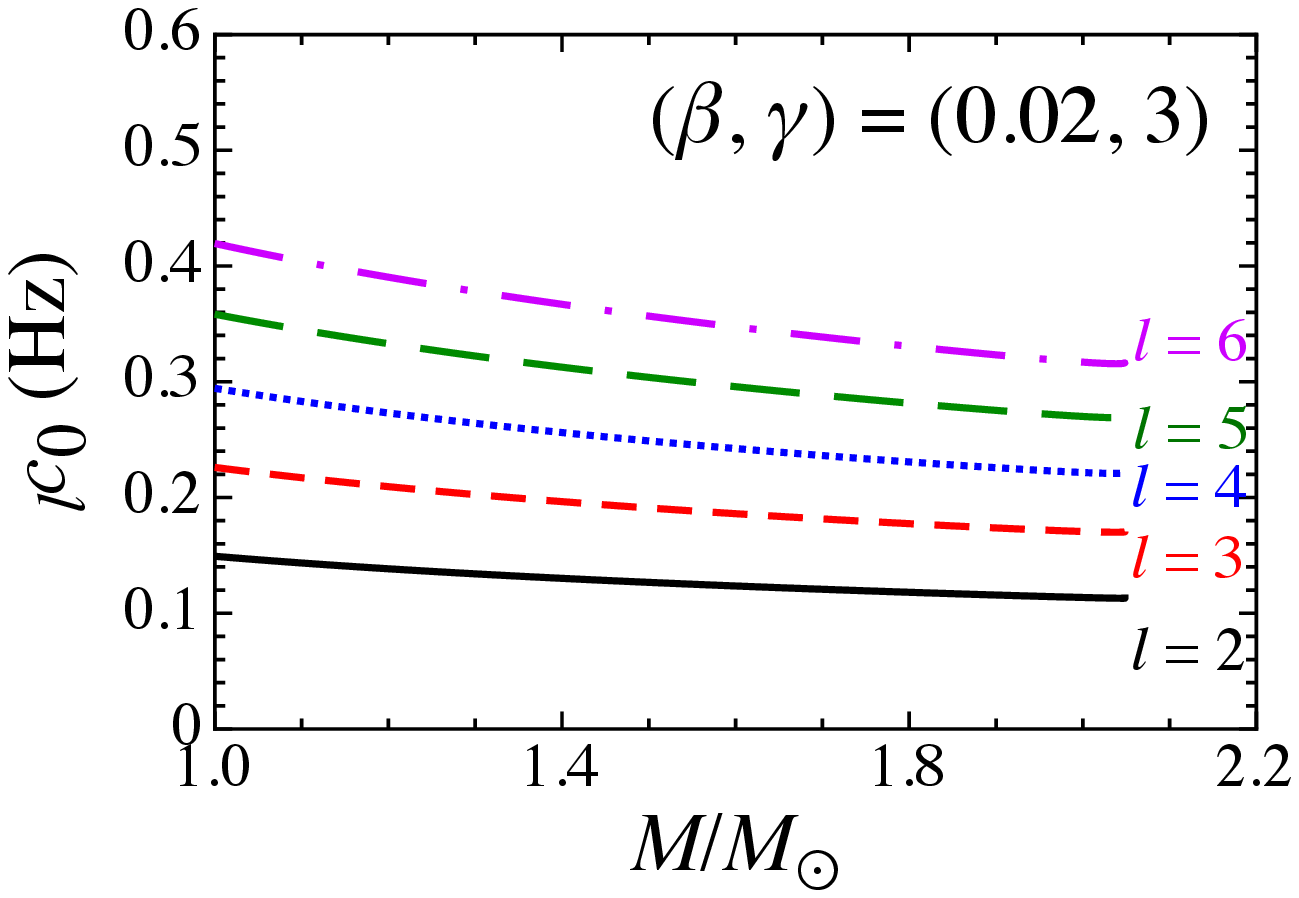} &
\includegraphics[scale=0.43]{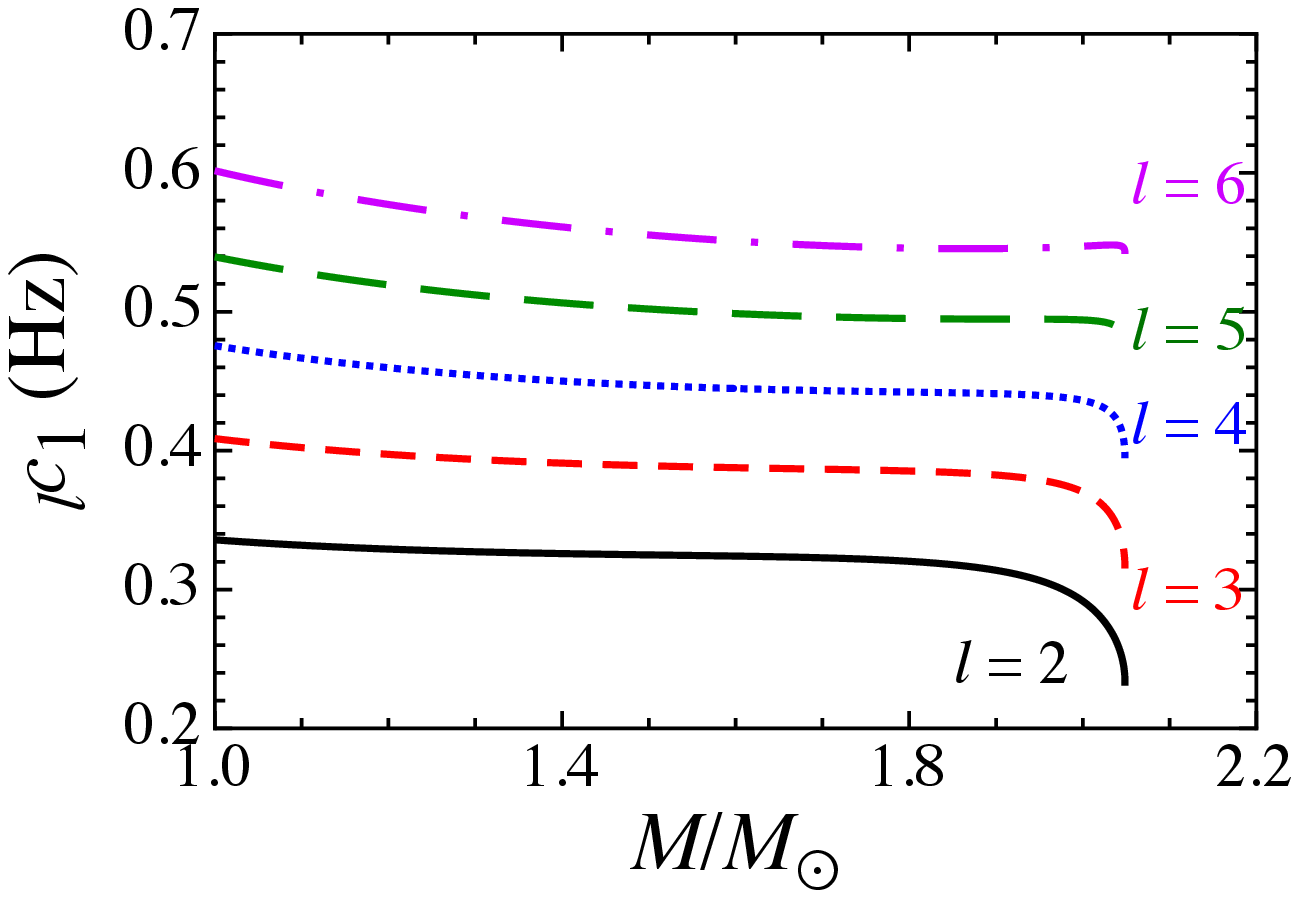} &
\includegraphics[scale=0.43]{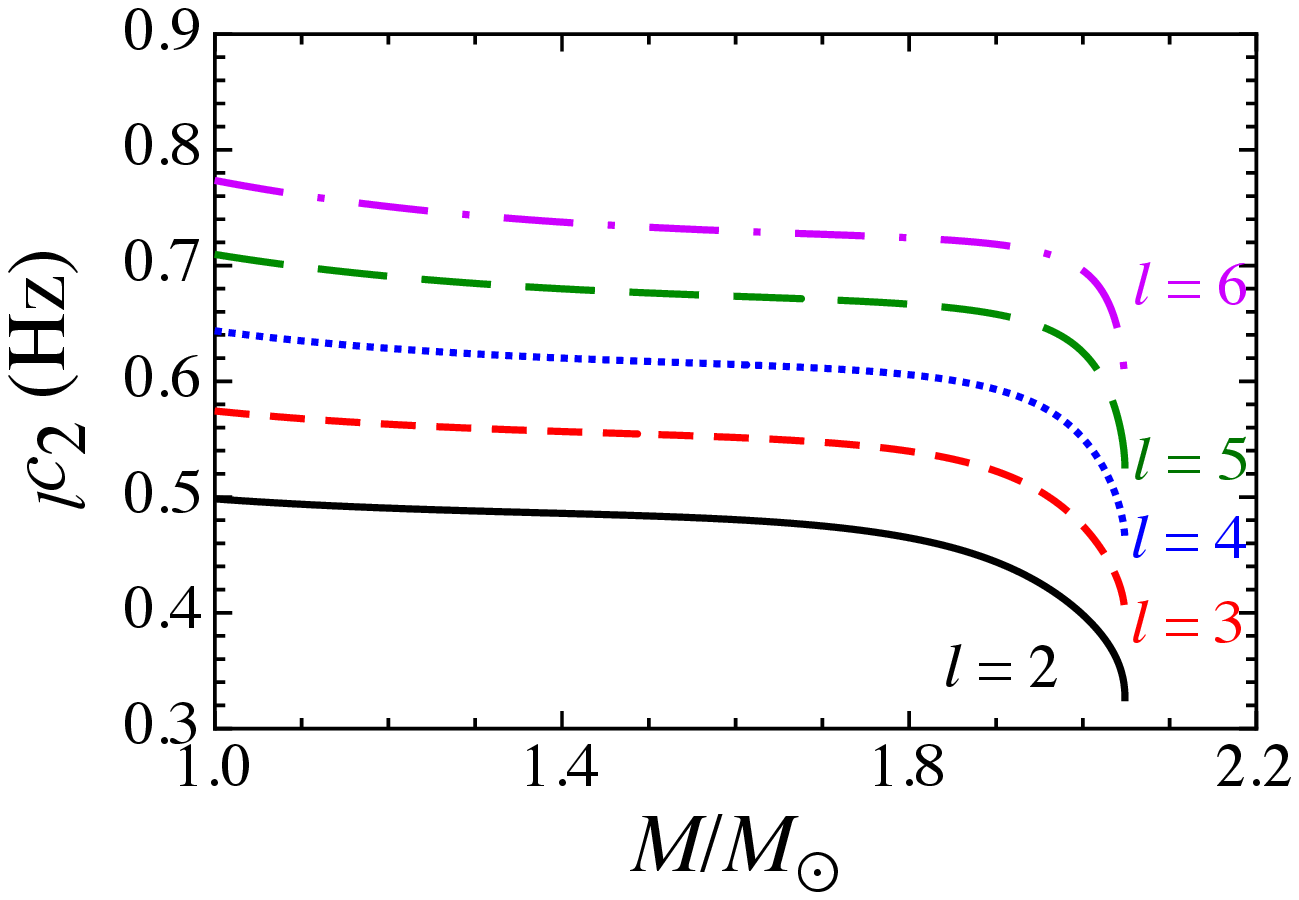}
\end{tabular}
\end{center}
\caption{
The coefficient in Eq. (\ref{eq:lan}) for the fundamental oscillations, ${}_\ell c_0$, in the left panel, the 1st overtones, ${}_\ell c_1$, in the middle panel, and the 2nd overtones, ${}_\ell c_2$, in the right panel for the $\ell=2$, $3$, $4$, $5$, and $6$ oscillations as a function of the stellar mass, where the strength distribution of magnetic fields is assumed to be $(\beta,\gamma) = (0.02, 3)$.
}
\label{fig:Mc-g3}
\end{figure*}

\begin{figure*}
\begin{center}
\begin{tabular}{ccc}
\includegraphics[scale=0.43]{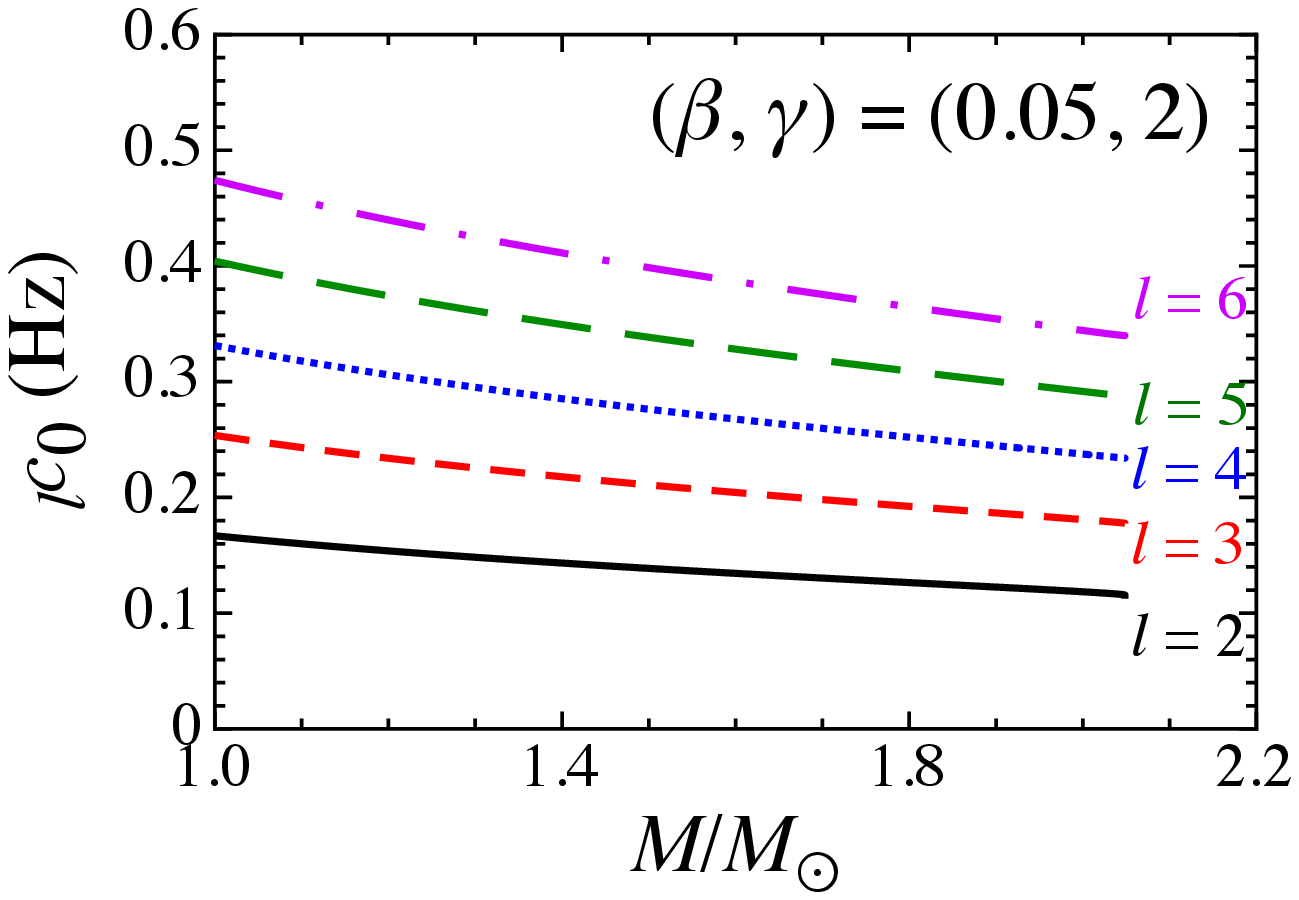} &
\includegraphics[scale=0.43]{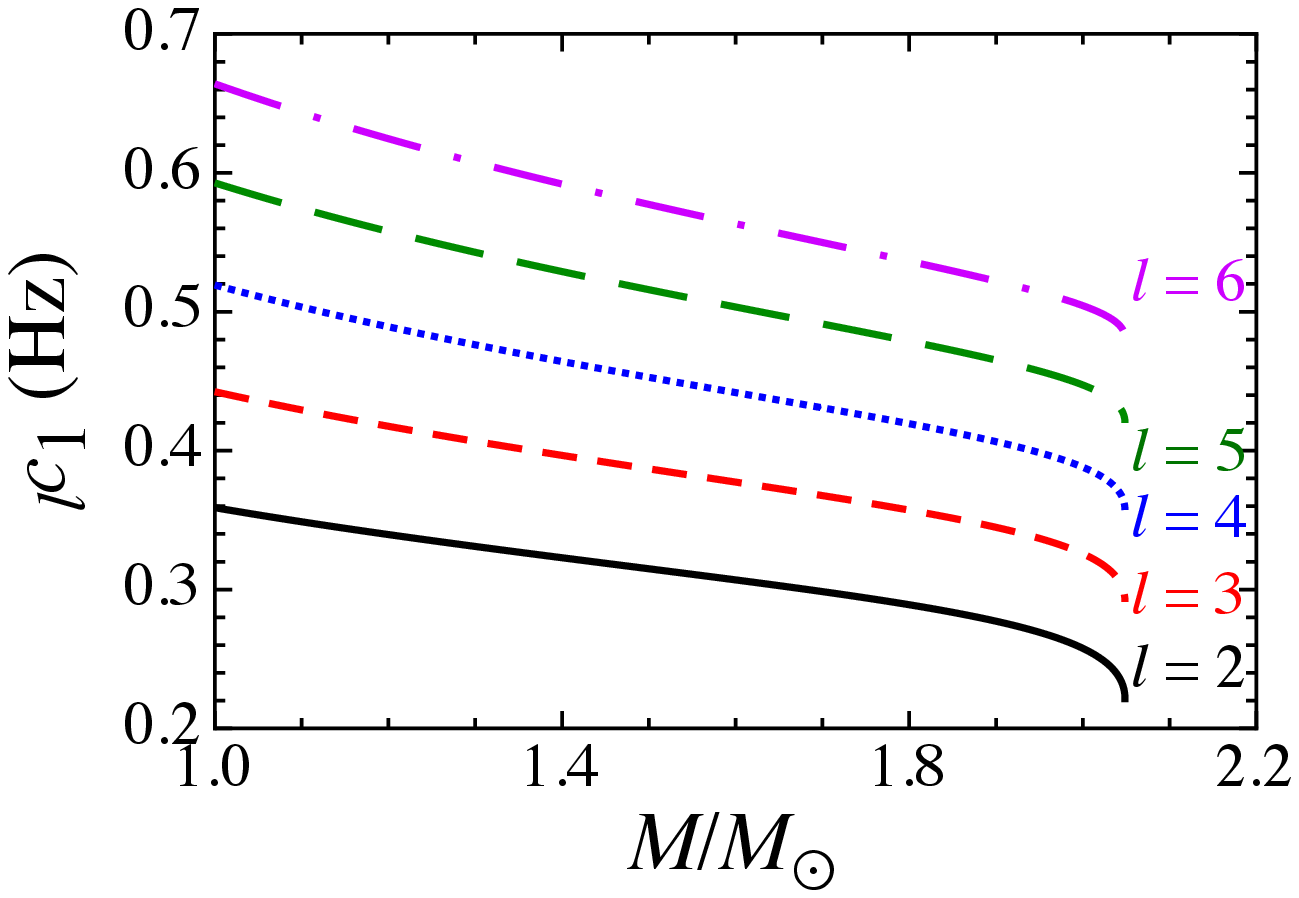} &
\includegraphics[scale=0.43]{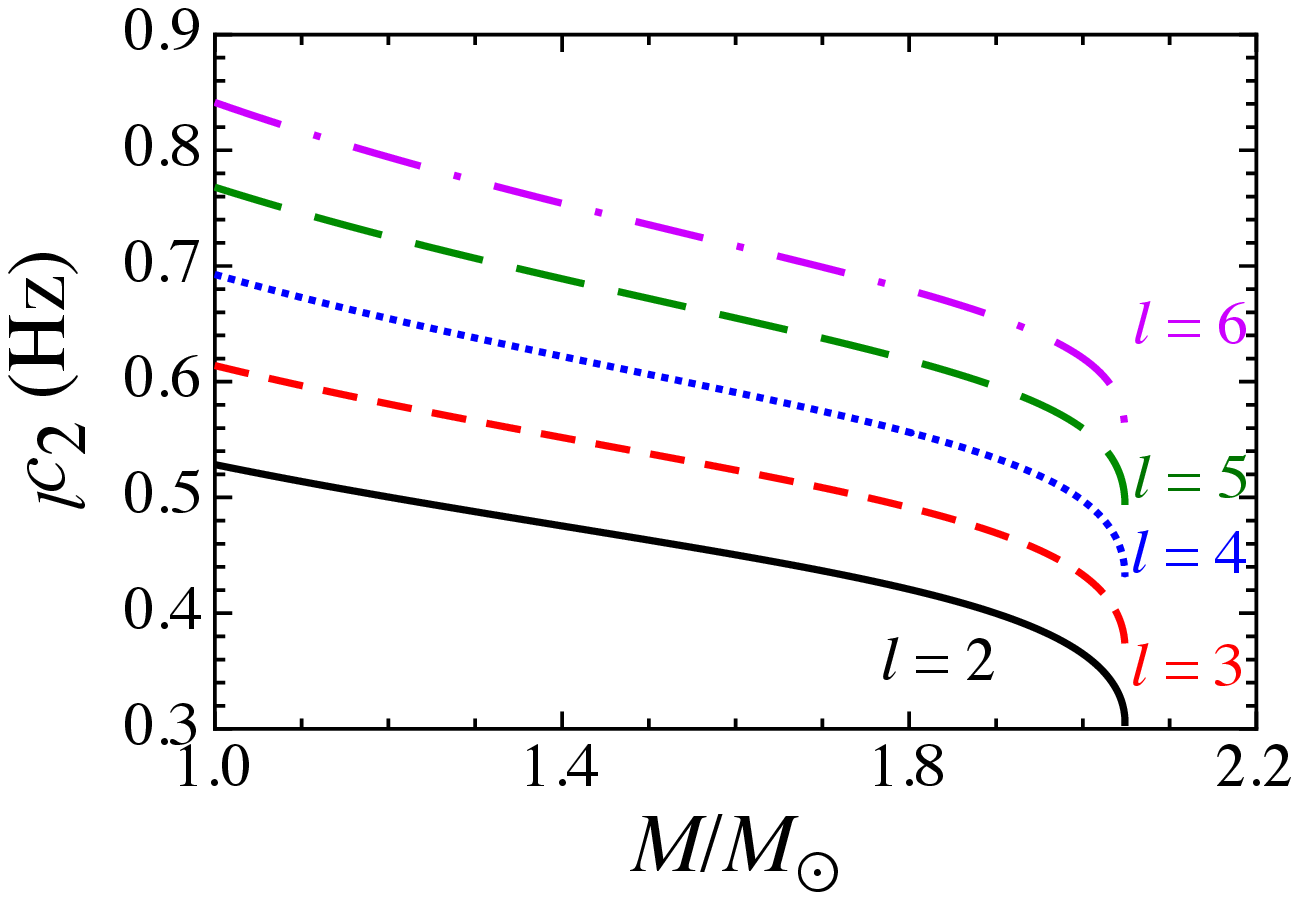}
\end{tabular}
\end{center}
\caption{
Same as Fig. \ref{fig:Mc-g3}, but for the strength distribution of magnetic fields with $(\beta,\gamma) = (0.05, 2)$.
}
\label{fig:Mc-g2}
\end{figure*}

The coefficient in Eq. (\ref{eq:lan}) for the fundamental oscillations, ${}_\ell c_0$, the 1st overtones, ${}_\ell c_1$, and the 2nd overtones, ${}_\ell c_2$ with respect to the $\ell=2$, $3$, $4$, $5$, and $6$ oscillations are shown in Fig. \ref{fig:Mc-g3} as a function of the stellar mass, where the strength distribution of magnetic fields is assumed to be $(\beta,\gamma) = (0.02, 3)$. Figure \ref{fig:Mc-g2} is the same as Fig. \ref{fig:Mc-g3}, but for $(\beta,\gamma) = (0.05, 2)$. We remark that the values of ${}_\ell c_n$ in Figs. \ref{fig:Mc-g3} and \ref{fig:Mc-g2} are equivalent to the corresponding frequencies, ${}_\ell a_n$, for the stellar model with ${\cal H}_{\rm surf}=10^{13}$ Gauss, as seen in Eq. (\ref{eq:lan}).

In both figures, the dependence of the coefficient in Eq. (\ref{eq:lan}) on the stellar mass is qualitatively very similar, because we adopt the specific strength distribution of magnetic fields such as Eq. (\ref{eq:He}). Nevertheless, the frequencies depend a little on the parameters of the magnetic distribution. That is, from Fig. \ref{fig:Mc-g3}, one observes that the coefficient in Eq. (\ref{eq:lan}) for the strength distribution with $(\beta,\gamma)=(0.02,3)$ becomes almost constant in the wide range of the stellar mass. On the other hand, from Fig. \ref{fig:Mc-g2}, one observes that the coefficient in Eq. (\ref{eq:lan}) for $(\beta,\gamma)=(0.05,2)$ decreases gradually as the stellar mass increases, compared with Fig. \ref{fig:Mc-g3}. This difference must come from the difference in the strength distribution of magnetic fields. Since the EOS of neutron star matter is still unknown, it is quite difficult to extract the information about the magnetic distributions inside the neutron star via the oscillation spectra. However, after one constrained the EOS for neutron stars via various future  observations, it may be possible to get an imprint of the strength distribution of magnetic fields by the observations of the oscillation spectra, with the help of the observation of the stellar mass.

Furthermore, we find a qualitative difference in the spectra of magnetic oscillations inside neutron stars with highly tangled magnetic fields compared to the situation with purely dipole magnetic fields. That is, the spectra in the case of purely dipole magnetic fields become continuum due to the difference of the lengths of magnetic field lines inside the star \cite{Sotani2008a,CBK2009,CSF2009,PL2013}, while those with highly tanged magnetic fields become discrete.

\section{Magnetic oscillations with crust elasticity}
\label{sec:VI}

\begin{figure}
\begin{center}
\includegraphics[scale=0.5]{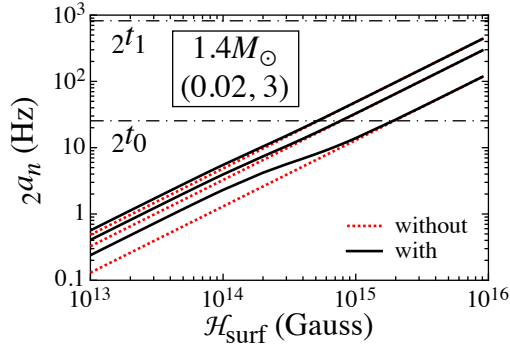} 
\end{center}
\caption{
The lowest three frequencies of the $\ell=2$ oscillations in the stellar model with $M=1.4M_\odot$ for the strength distribution with $(\beta,\gamma)=(0.02,3)$, shown as a function of ${\cal H}_{\rm surf}$, where the solid and dotted lines correspond to the results with and without crust elasticity. The horizontal dot-dashed lines denote the frequencies of the $\ell=2$ crustal torsional oscillations without magnetic fields.
}
\label{fig:wHs-M14}
\end{figure}

Now, we consider the magnetic oscillations with the effect of crust elasticity. In Fig. \ref{fig:wHs-M14}, we show the lowest three frequencies of the $\ell=2$ oscillations with the solid lines for the stellar model with $M=1.4M_\odot$ and $(\beta,\gamma)=(0.02,3)$, as a function of ${\cal H}_{\rm surf}$, where the results without the crust elasticity are also shown with the dotted lines for reference. From this figure, one can observe that the effect of crust elasticity on the frequencies would disappear for the stellar model with stronger magnetic fields. This behavior is the same as in the case with the pure dipole magnetic fields \cite{CK2011,GCFMS2012}. This is why the Alfv\'{e}n velocity, defined as $v_{\rm A}\equiv {\cal H}/\sqrt{\varepsilon}$, dominates inside a star with a magnetic field stronger than a critical strength, where the shear velocity characterized by the crust elasticity, $v_{\rm s}\equiv (\mu/\varepsilon)^{1/2}$, becomes relatively negligible. The typical value of the critical strength of the magnetic fields is considered so that the Alfv\'{e}n velocity becomes equivalent to the shear velocity at the crust basis ($r=R_{\rm c}$) \cite{Sotani2007}. With the shear modulus given by Eq. (\ref{eq:shear}) and the SLy4 EOS adopted in this paper, one can determine the typical value of the critical strength at the crust basis to be ${\cal H}=1.52\times 10^{15}$ Gauss, which leads to the strength at the stellar surface ${\cal H}_{\rm surf}=1.50\times 10^{15}$ and $1.44\times 10^{15}$ Gauss for $(\beta,\gamma)=(0.02,3)$ and $(0.05,2)$, respectively. In fact, the difference between the frequencies with and without crust elasticity can disappear for the magnetic fields stronger than ${\cal H}_{\rm surf}\simeq 1.50\times 10^{15}$ in Fig. \ref{fig:wHs-M14}. Here, we show only the case with $(\beta,\gamma)=(0.02,3)$, but the results with $(\beta,\gamma)=(0.05,2)$
are quite similar to Fig. \ref{fig:wHs-M14}.

\begin{figure}
\begin{center}
\includegraphics[scale=0.5]{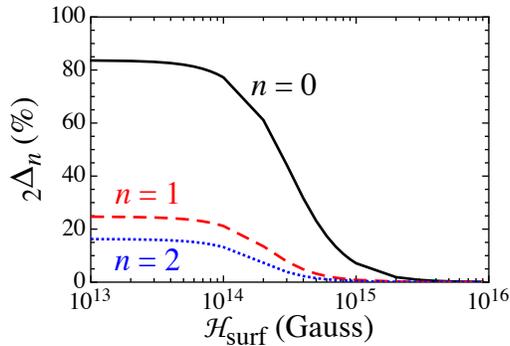} 
\end{center}
\caption{
Relative deviation between the $\ell=2$ frequencies with and without crust elasticity as a function of ${\cal H}_{\rm surf}$, where $n=0$, $1$, and $2$ correspond to the fundamental oscillations, 1st, and 2nd overtones, respectively. The relative deviation is defined by Eq. (\ref{eq:lDnw}).
}
\label{fig:wHs-M14d}
\end{figure}

On the other hand, we find that the frequencies for the stellar model with weak magnetic fields, which deviate from those without crust elasticity, become completely different from the frequencies of the crustal torsional oscillations without magnetic fields. In Fig. \ref{fig:wHs-M14}, the horizontal dot-dashed lines correspond to the frequencies of the $\ell=2$ crustal torsional oscillations without magnetic fields, which are denoted by ${}_\ell t_n$ for the frequencies of the $\ell$th oscillations with the number of the radial nodes, $n$. That is, ${}_2 t_0$ and ${}_2 t_1$ are the frequencies of the fundamental oscillations and 1st overtone for $\ell=2$ torsional oscillations, which become ${}_2 t_0=25.5$ Hz and ${}_2 t_1=820.8$ Hz for the stellar model with $M=1.4M_\odot$. Such behavior of the frequencies with respect to the strength of the magnetic fields must be a feature owing to the highly tangled magnetic fields inside the star. Actually, it was shown that, for the stellar models with purely dipole magnetic fields, the oscillations excited in the vicinity of the stellar surface become the crustal torsional oscillations, if the strength of the magnetic fields is weak \cite{GCFMS2012}.

In order to clearly see the difference in the frequencies due to the existence of the crust elasticity, we also calculate the relative deviation between the $\ell=2$ frequencies with and without crust elasticity, which is shown in Fig. \ref{fig:wHs-M14d} as a function of ${\cal H}_{\rm surf}$. Here, the relative deviation is calculated by
\begin{equation}
  {}_\ell \Delta_n = \frac{{}_\ell a_n^{\rm w} - {}_\ell a_n^{\rm wo}}{{}_\ell a_n^{\rm wo}},  \label{eq:lDnw}
\end{equation}
where ${}_\ell a_n^{\rm w}$ and ${}_\ell a_n^{\rm wo}$ denote the frequencies with and without crust elasticity, respectively, for the $\ell$th magnetic oscillations with the number of radial nodes, $n$. In this figure, the solid, broken, and dotted lines correspond to the relative deviation for the fundamental oscillations ($n=0$), the 1st overtones ($n=1$), and the 2nd overtones ($n=2$), respectively. From this figure, we find that the effect of the crust elasticity appears stronger in the oscillations with smaller radial nodes. Additionally, we observe that the relative deviation from the frequencies without crust elasticity seems to be almost constant if the strength of the magnetic fields is significantly weak. In other words, the relative deviation, ${}_2 \Delta_n$, depends on the magnetic field strength in the range between $\sim\tilde{\cal H}/10$ and $\sim\tilde{\cal H}$, where $\tilde{\cal H}$ denotes the typical critical field strength at the stellar surface so that the shear velocity becomes equivalent to the Alfv\'{e}n velocity at the crust basis, i.e., $\tilde{\cal H}=1.50\times 10^{15}$ Gauss for the stellar model constructed with the SLy4 EOS and $(\beta,\gamma)=(0.02,3)$, as mentioned before.

\begin{figure*}
\begin{center}
\begin{tabular}{cc}
\includegraphics[scale=0.5]{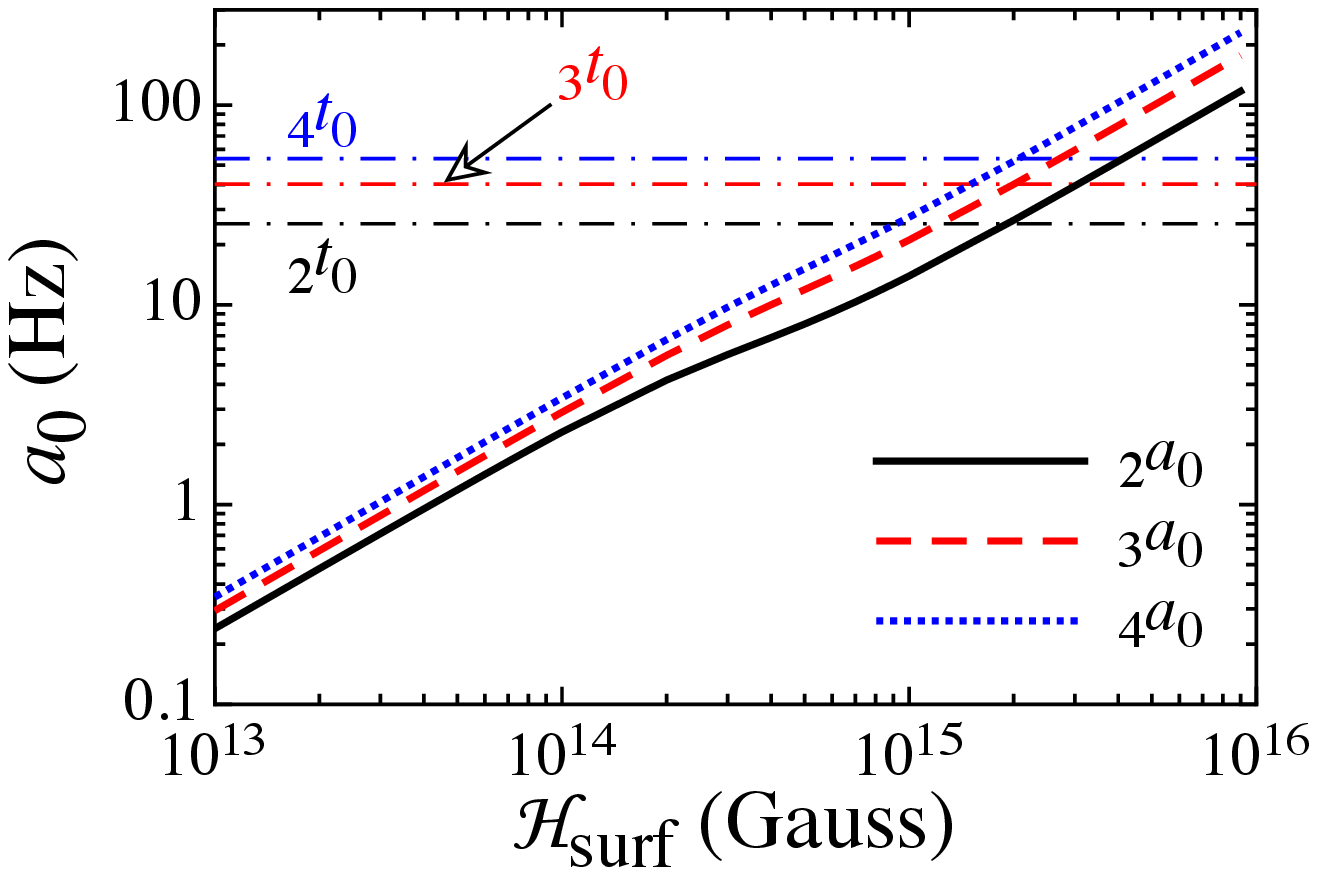} &
\includegraphics[scale=0.5]{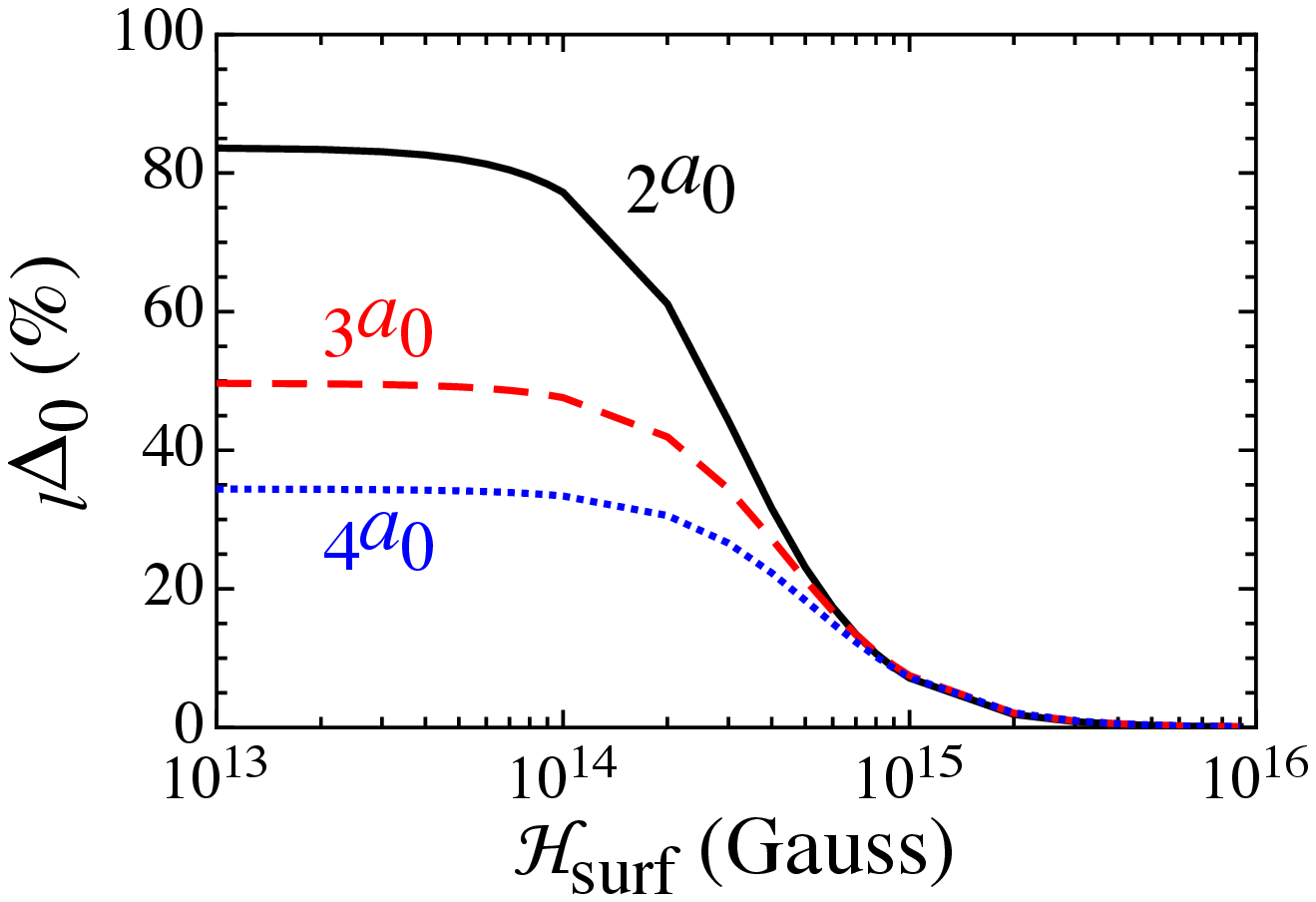}
\end{tabular}
\end{center}
\caption{
In the left panel, the frequencies of the $\ell=2$, $3$, and $4$ fundamental magnetic oscillations, ${}_2a_0$, ${}_3a_0$, and ${}_4a_0$, are shown as a function of ${\cal H}_{\rm surf}$ for the stellar model with $M=1.4M_\odot$ and $(\beta,\gamma)=(0.02,3)$, with the frequencies of the $\ell=2$, $3$, and $4$ fundamental torsional oscillations, ${}_2t_0$, ${}_3t_0$, and ${}_4t_0$ for reference. In the right panel, the relative deviation between the frequencies with and without crust elasticity is shown as a function of ${\cal H}_{\rm surf}$.
}
\label{fig:wHs-a0}
\end{figure*}

In the left panel of Fig. \ref{fig:wHs-a0}, we show the frequencies of the $\ell=2$, $3$, and 4 fundamental magnetic oscillations, ${}_2a_0$, ${}_3a_0$, and ${}_4a_0$, as a function of ${\cal H}_{\rm surf}$ for the stellar model with $M=1.4M_\odot$ and $(\beta,\gamma)=(0.02,3)$, while in the right panel of Fig. \ref{fig:wHs-a0} we show the relative deviation between the frequencies with and without crust elasticity as a function of ${\cal H}_{\rm surf}$. As with the frequencies of $\ell=2$ magnetic oscillations shown in Fig. \ref{fig:wHs-M14}, the effect of the crust elasticity can disappear for the stellar model with strong magnetic fields.  For reference, we also show the $\ell=2$, $3$, and $4$ fundamental torsional oscillations, ${}_2t_0$, ${}_3t_0$, and ${}_4t_0$, in the left panel of Fig. \ref{fig:wHs-a0}, from which one observes that the frequencies of the $\ell$th magnetic oscillations for the stellar model with weak magnetic fields are completely different from those of the torsional oscillations. Meanwhile, from the right panel, we find that the effect of crust elasticity becomes  stronger for the magnetic oscillations with lower $\ell$ for the stellar model with weak magnetic fields, and also that ${}_\ell \Delta_0$ depends strongly on the magnetic field strength in the range between $\sim\tilde{\cal H}/10$ and $\sim\tilde{\cal H}$ again as in Fig. \ref{fig:wHs-M14d}.

\begin{figure*}
\begin{center}
\begin{tabular}{ccc}
\includegraphics[scale=0.43]{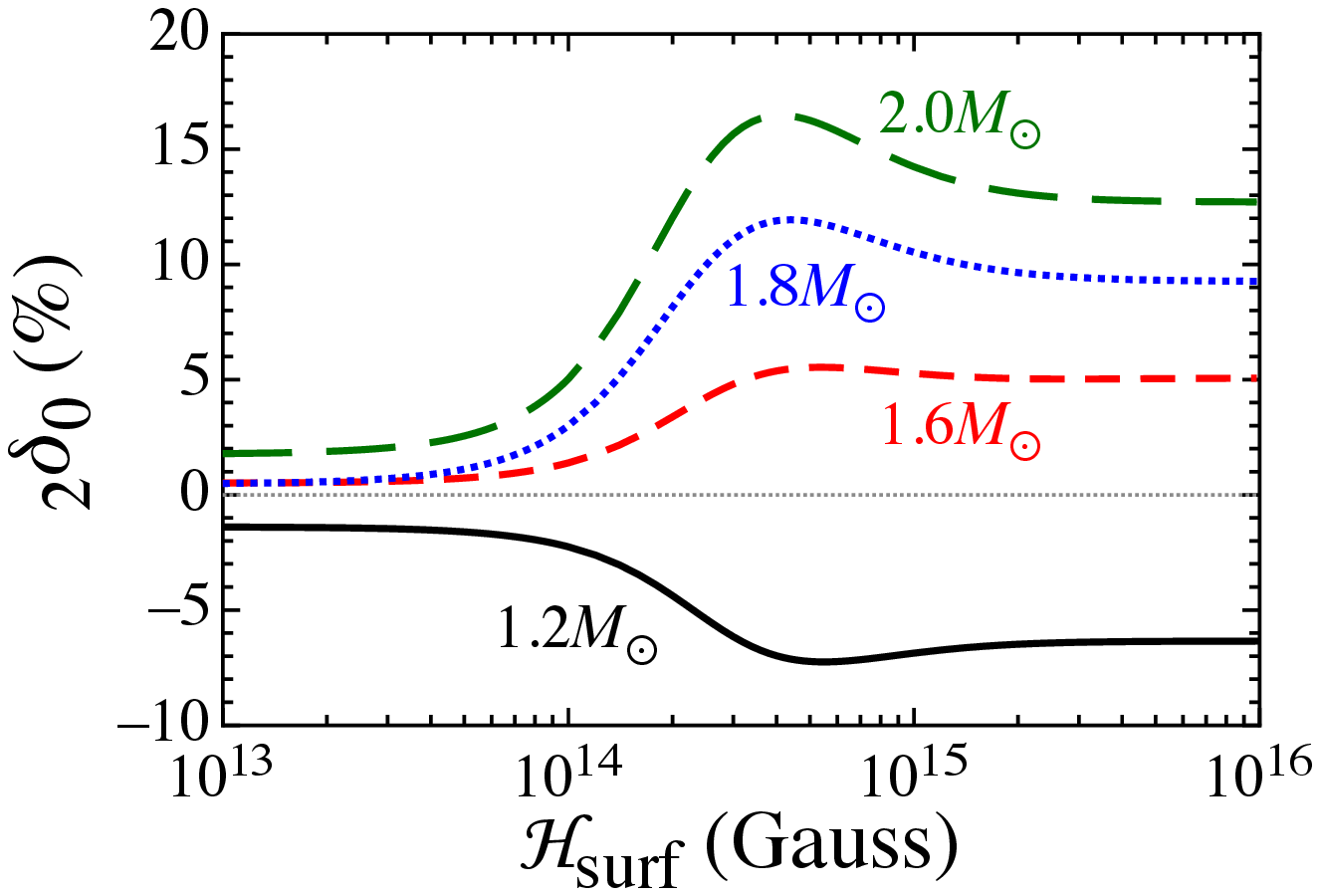} &
\includegraphics[scale=0.43]{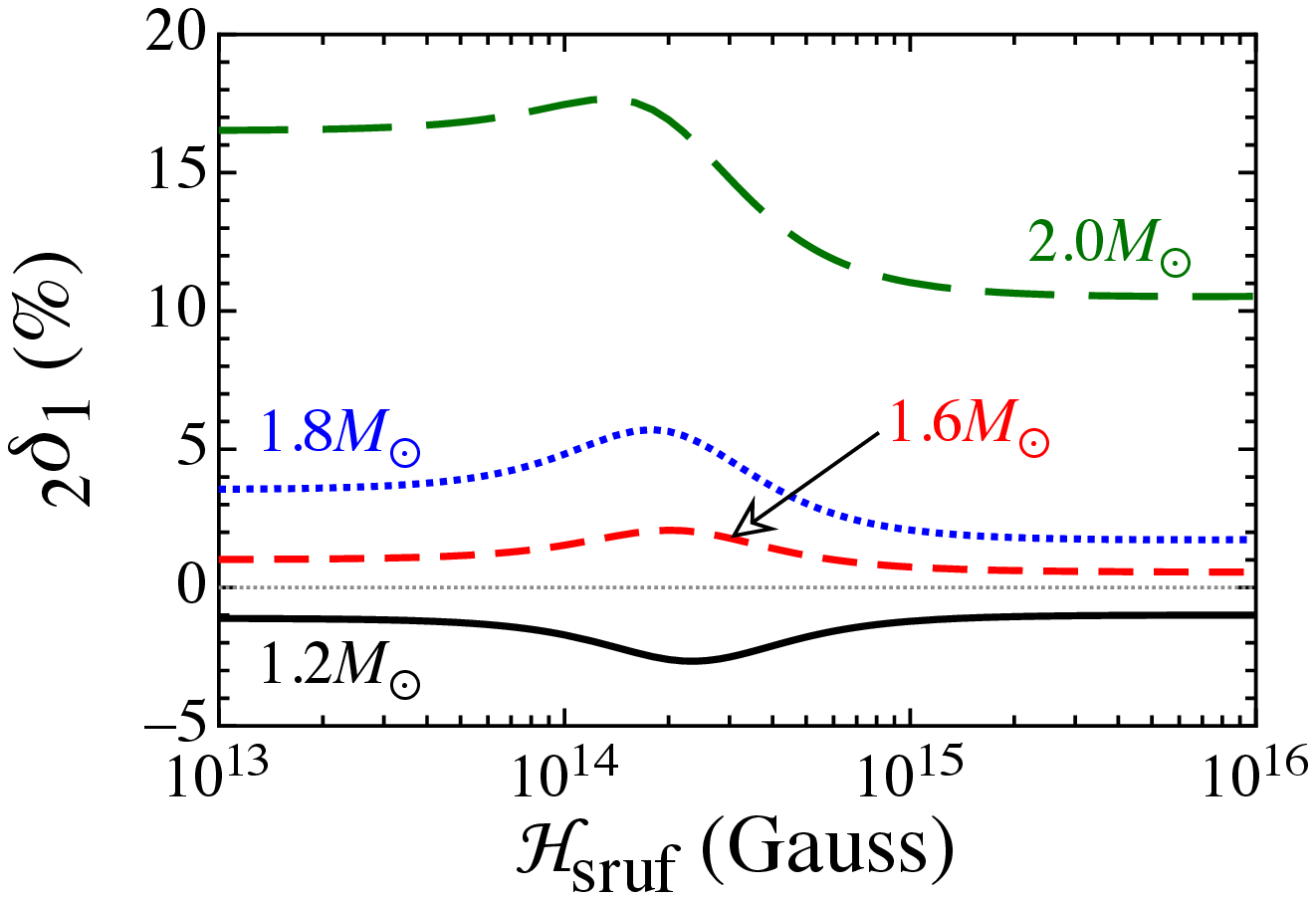} &
\includegraphics[scale=0.43]{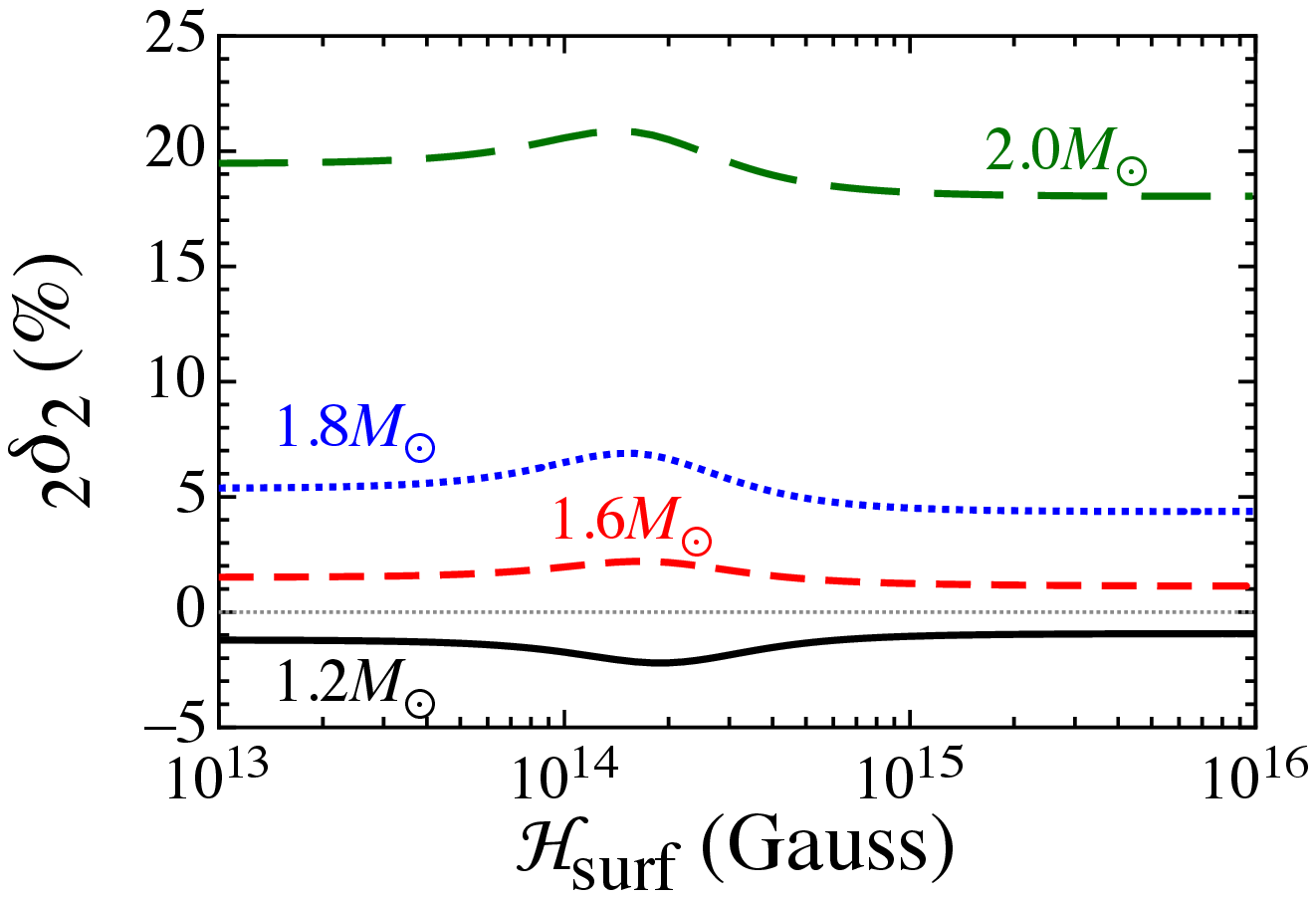}
\end{tabular}
\end{center}
\caption{
Relative deviation of the frequencies of ${}_2a_n$ with $n=0$, $1$, and $2$ for the various stellar models from those for the stellar model with $M=1.4M_\odot$, as a function of ${\cal H}_{\rm surf}$, where the relative deviation is defined by Eq. (\ref{eq:ldn-M}) and the parameters in the strength distribution of magnetic fields are $(\beta,\gamma)=(0.02,3)$.
}
\label{fig:2d-Hs}
\end{figure*}

\begin{figure*}
\begin{center}
\begin{tabular}{ccc}
\includegraphics[scale=0.43]{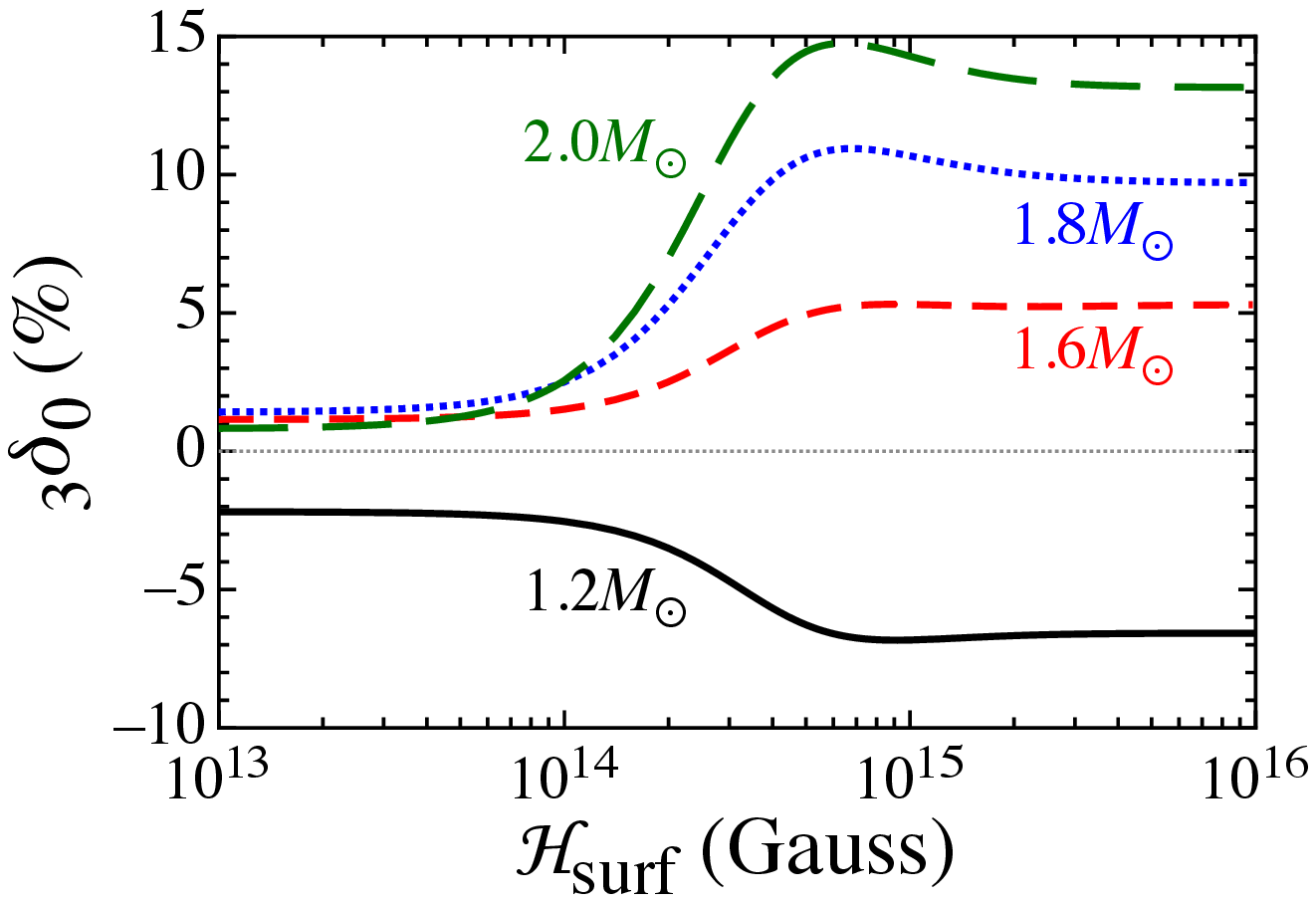} &
\includegraphics[scale=0.43]{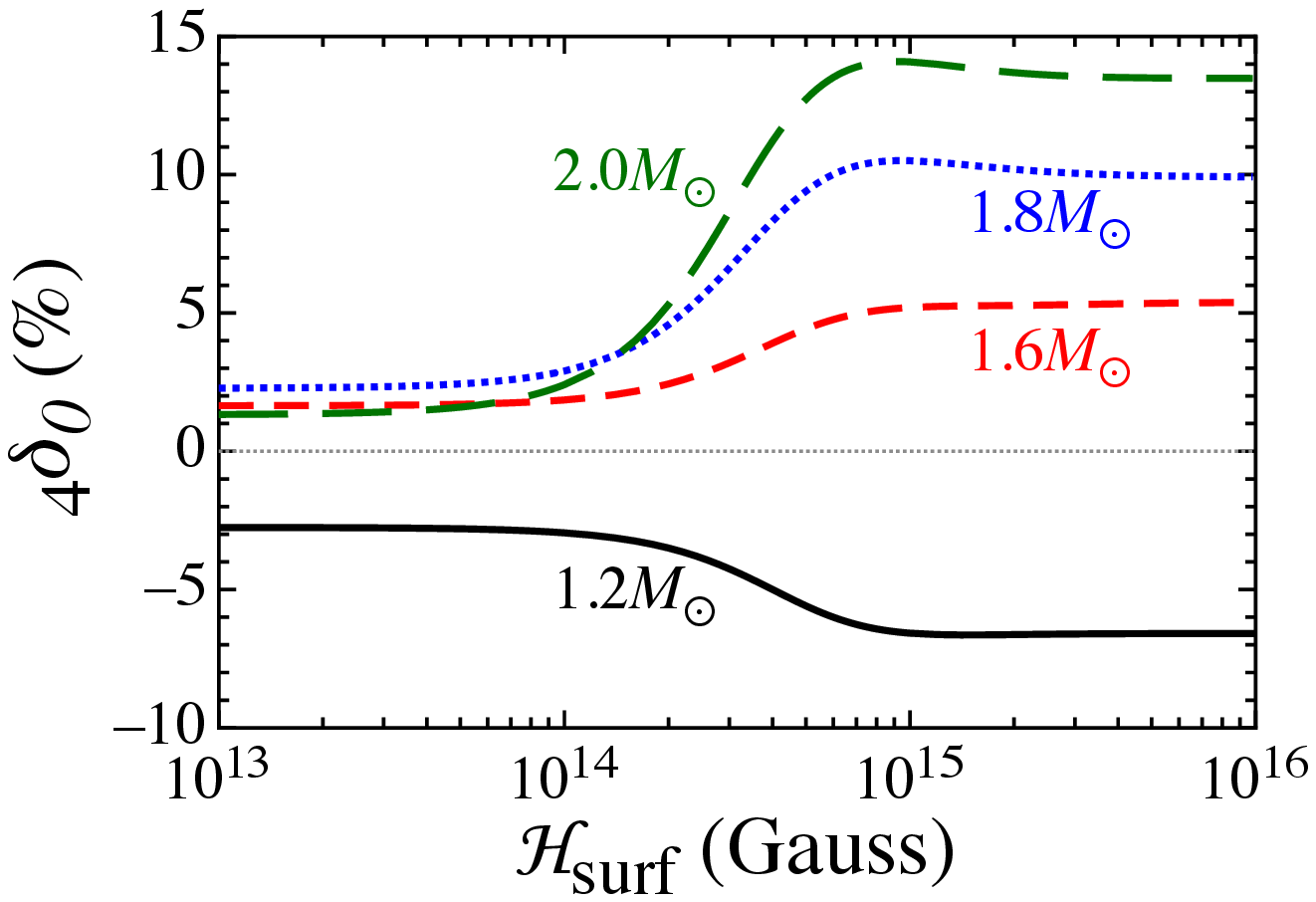} &
\includegraphics[scale=0.43]{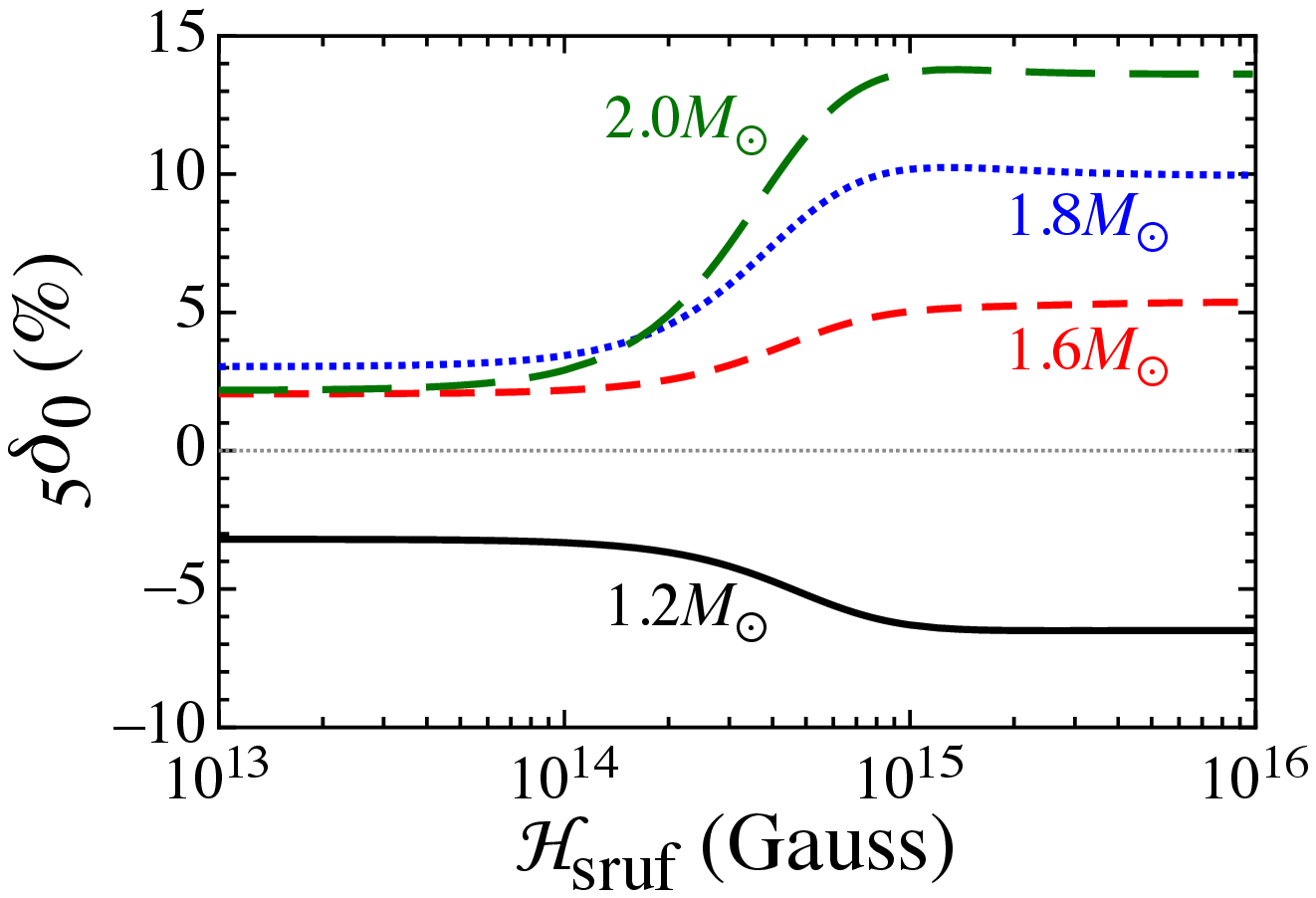}
\end{tabular}
\end{center}
\caption{
Same as Fig. \ref{fig:2d-Hs}, but for the fundamental oscillations with $\ell=3$, $4$, and $5$.
}
\label{fig:d0-Hs}
\end{figure*}

Moreover, in order to see how the frequencies of magnetic oscillations could be shifted for the different stellar models, we focus on the relative deviation of the frequencies from those for the stellar model with $M=1.4M_\odot$. So, the relative deviation is evaluated by
\begin{equation}
   {}_\ell \delta_n = \frac{{}_\ell a_n^{14} - {}_\ell a_n^M}{{}_\ell a_n^{14}},   \label{eq:ldn-M}
\end{equation}
where ${}_\ell a_n^{14}$ and ${}_\ell a_n^M$ denote the frequencies of the $\ell$th magnetic oscillations with the number of the radial nodes, $n$, for the neutron star models with $M=1.4M_\odot$ and with the stellar mass, $M$, respectively. Then, using the calculated frequencies of magnetic oscillations for the various stellar models, we show the values of ${}_2\delta_n$ for $n=0$, $1$, and $2$ in Fig. \ref{fig:2d-Hs}, and ${}_\ell \delta_0$ for $\ell=3$, $4$, and $5$ in Fig. \ref{fig:d0-Hs} as a function of ${\cal H}_{\rm surf}$. From the both figures, one can observe that the relative deviation, ${}_\ell \delta_n$, depends strongly on the magnetic field strength in the range of ${\cal H}_{\rm surf}\sim\tilde{\cal H}/10-\tilde{\cal H}$, while ${}_\ell \delta_n$ becomes almost constant in the other region of the field strength. The behaviors of the relative deviation for the $\ell$-th fundamental magnetic oscillations, ${}_\ell \delta_0$, are very similar to each other at least up to $\ell=5$. In particular, we find that the frequencies of the $\ell$th fundamental oscillations for the weak field strength are less dependent on the stellar mass, which is less than a few percent difference from the frequency expected for the stellar model with $M=1.4M_\odot$. In addition, the frequencies of the $\ell$th fundamental magnetic oscillations depend on the stellar mass for the strong magnetic fields, whose dependence should be similar to Fig. \ref{fig:Mc-g3} for the magnetic oscillations without the effect of crust elasticity, because the magnetic oscillations under the strong magnetic fields are almost independent of the existence of the crust elasticity. On the other hand, the frequencies of the overtones of magnetic oscillations depend on the stellar mass not only in the strong field regime but also in the weak field regime, as shown in Fig. \ref{fig:2d-Hs}.

\begin{figure}
\begin{center}
\includegraphics[scale=0.5]{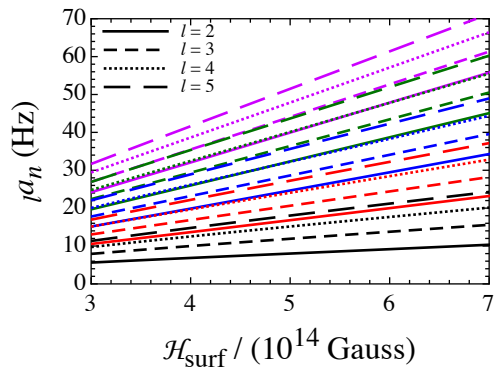} 
\end{center}
\caption{
Various eigenfrequencies, ${}_\ell a_n$ for the stellar model with $M=1.4M_\odot$ and $(\beta,\gamma)=(0.02,3)$ in the short range of ${\cal H}_{\rm surf}$, where we show the frequencies for $\ell=2$, $3$, $4$, and $5$ and $n=0$, $1$, $2$, $3$, and $4$. That is, the frequencies for ${\cal H}_{\rm surf}=7\times 10^{14}$ Gauss correspond to ${}_2a_0$, ${}_3a_0$, ${}_4a_0$, ${}_2a_1$, ${}_5a_0$, ${}_3a_1$, ${}_4a_1$, ${}_2a_2$, ${}_5a_1$, ${}_3a_2$, ${}_4a_2$, ${}_2a_3$, ${}_5a_2$, ${}_3a_3$, ${}_4a_3$, ${}_2a_4$, ${}_5a_3$, ${}_3a_4$, ${}_4a_4$, and ${}_5a_4$ in order from bottom to top.
}
\label{fig:lan-M14}
\end{figure}

\begin{figure}
\begin{center}
\includegraphics[scale=0.5]{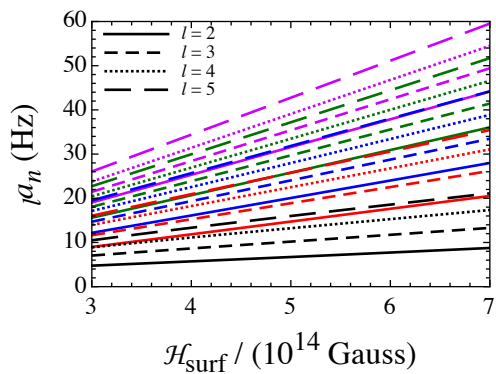} 
\end{center}
\caption{
Same as Fig. \ref{fig:lan-M14}, but for the stellar model with $M=2.0M_\odot$, where the frequencies for ${\cal H}_{\rm surf}=7\times 10^{14}$ Gauss correspond to ${}_2a_0$, ${}_3a_0$, ${}_4a_0$, ${}_2a_1$, ${}_5a_0$, ${}_3a_1$, ${}_2a_2$, ${}_4a_1$, ${}_3a_2$, ${}_5a_1$, ${}_2a_3$, ${}_4a_2$, ${}_3a_3$, ${}_2a_4$, ${}_5a_2$, ${}_4a_3$, ${}_3a_4$, ${}_5a_3$, ${}_4a_4$, and ${}_5a_4$ in order from bottom to top.
}
\label{fig:lan-M20}
\end{figure}

Due to such a complex dependence of the frequencies on the stellar mass, the spectra from the neutron stars with highly tangled magnetic fields also become complex especially in the range of ${\cal H}_{\rm surf}\sim \tilde{\cal H}/10-\tilde{\cal H}$. As an example, we show that the frequencies, ${}_\ell a_n$, for $\ell=2$, $3$, $4$, and $5$, and $n=0$, $1$, $2$, $3$, and $4$ expected in the neutron stars with $M=1.4M_\odot$ in Fig. \ref{fig:lan-M14}, and with $M=2.0M_\odot$ in Fig. \ref{fig:lan-M20}, as a function of ${\cal H}_{\rm surf}$ in the range of ${\cal H}_{\rm surf}=(3-7)\times 10^{14}$ Gauss. Comparing the both figures, one can observe that the order of the eigenfrequencies depends on the stellar mass. That is, focusing on the field strength of ${\cal H}_{\rm surf}=7\times 10^{14}$ Gauss, the frequencies for the stellar model with $M=1.4M_\odot$ correspond to ${}_2a_0$, ${}_3a_0$, ${}_4a_0$, ${}_2a_1$, ${}_5a_0$, ${}_3a_1$, ${}_4a_1$, ${}_2a_2$, ${}_5a_1$, ${}_3a_2$, ${}_4a_2$, ${}_2a_3$, ${}_5a_2$, ${}_3a_3$, ${}_4a_3$, ${}_2a_4$, ${}_5a_3$, ${}_3a_4$, ${}_4a_4$, and ${}_5a_4$ in order from bottom to top, while those with $M=2.0M_\odot$ are ${}_2a_0$, ${}_3a_0$, ${}_4a_0$, ${}_2a_1$, ${}_5a_0$, ${}_3a_1$, ${}_2a_2$, ${}_4a_1$, ${}_3a_2$, ${}_5a_1$, ${}_2a_3$, ${}_4a_2$, ${}_3a_3$, ${}_2a_4$, ${}_5a_2$, ${}_4a_3$, ${}_3a_4$, ${}_5a_3$, ${}_4a_4$, and ${}_5a_4$ in order from bottom to top. In any case, there is the forbidden region in the spectra, which corresponds to the region below the line of ${}_2 a_0$. We also emphasize that, unlike the pure crust torsional oscillations, one can expect many of the eigenfrequencies in the magnetic oscillations. This is because the overtone frequencies of the crustal torsional oscillations become much higher than the frequencies of the fundamental oscillations, while those of the magnetic oscillations become in the same order as for the fundamental oscillations.

\section{Conclusion}
\label{sec:V}

In this paper, we systematically examine the frequencies of the magnetic oscillations in neutron stars with highly tangled magnetic fields, focusing on the axial type oscillations. For this purpose, we derive the perturbation equations describing such oscillations by combining the linearized equation of motion and induction equation. To derive this perturbation equation, we assume that the strength of the global magnetic structure is much smaller than the tangled field strength, and that the magnetic fields are tangled with a length scale smaller than the wavelength of the magnetic oscillations considered in this paper. Then, we calculate the frequencies of magnetic oscillations with and without crust elasticity, adopting the phenomenological strength distribution of magnetic fields.

We confirm that the frequencies of magnetic oscillations without crust elasticity are exactly proportional to the field strength, as expected. The frequencies decrease as the stellar mass increases, which also depend on the strength distribution of magnetic fields. On the other hand, the spectra of the magnetic oscillations with crust elasticity become more complicated, where we could not observe the pure crustal torsional oscillations even for the weak magnetic fields. For discussing the spectra, we show the importance of the critical field strength at the stellar surface, $\tilde{\cal H}$, determined in such a way that the shear velocity is equivalent to the Alfv\'{e}n velocity at the crust basis. In fact, we find that, independently of the stellar mass, the frequencies are almost proportional to the strength of magnetic fields except for the range from $\sim\tilde{\cal H}/10$ up to $\sim\tilde{\cal H}$. The effect of the crust elasticity can be seen more strongly in the fundamental oscillations with lower harmonics index $\ell$. Additionally, we show that the fundamental oscillations are less dependent on the stellar mass for the weak magnetic fields, while the overtones are more sensitive to the stellar mass not only in the weak but also the  strong field regimes. Furthermore, we find that the spectra of the magnetic oscillations in the neutron stars with highly tangled magnetic fields are discrete. This is completely a different spectrum property from the case for the stellar model with pure dipole magnetic fields, which leads to the continuum spectra.

In this paper, we do not take into account the contributions from the global magnetic structure, which should play an important role for considering the cold neutron stars. On the other hand, Link and van Eysden studied the full range between a purely ordered field and a purely tangled field, even though their stellar models are quite simple \cite{LvE2015}, which could give a more reliable qualitative picture than the limiting case of a purely tangled field, at least for the cold neutron stars. At some point, we will examine the oscillation spectra on the stellar models, including the contribution of the global magnetic structure. In any event, we have figured out that, unlike the crustal torsional oscillations, one can observe many magnetic oscillations in the spectra, which may be detected after the violent phenomena breaking the global magnetic structure.

\acknowledgments
This work was supported in part by Grant-in-Aid for Young Scientists (B) through Grant No. 26800133 provided by JSPS and by Grants-in-Aid for Scientific Research on Innovative Areas through Grant No. 15H00843 provided by MEXT.



\end{document}